\documentclass[11pt,nofootinbib]{article}
\pdfoutput=1
\usepackage{jheppub}
\usepackage{graphicx}
\usepackage{slashed}
\usepackage{subfig}
\usepackage[normalem]{ulem}
\usepackage{wasysym}
\usepackage{cancel}
\usepackage[utf8]{inputenc}

\newcommand\beq{\begin{equation}}
\newcommand\eeq{\end{equation}}

\newcommand\nn{\nonumber}

\title{Dark Matter-Neutrino Interaction in Light of Collider and Neutrino Telescope Data
}
\author[a]{Reinard Primulando,}
\author[b]{Patipan Uttayarat}

\affiliation[a]{Center for Theoretical Physics, Department of Physics, Parahyangan Catholic University, Jl. Ciumbuleuit 94, Bandung 40141, Indonesia}
\affiliation[b]{Department of Physics, Srinakharinwirot University, 114 Sukhumvit 23rd Rd., Wattana, Bangkok 10110, Thailand}

\emailAdd{rprimulando@unpar.ac.id}
\emailAdd{patipan@g.swu.ac.th}

\abstract{ 
We study the DM-neutrino interaction in the framework of simplified model. 
The phenomenology of such an interaction are derived. 
We also investigate the bound on DM-neutrino interaction from the LHC and neutrino telescopes. 
We find that for the case of a scalar dark matter, the LHC gives a stronger bound on dark matter annihilation cross-section than the neutrino telescopes.
However, for the fermionic dark matter case the neutrino telescopes bounds are more stringent for dark matter mass, $\gtrsim 200$ MeV. 
In the case of lower DM mass, the neutrino telescopes provide better bounds for a light mediator, while the collider bounds are better for a heavy mediator. 
Possible UV completions of the simplified model are briefly discussed.   
}

\begin{document}
\maketitle

\flushbottom
\section{Introduction}
The LHC has played an important role in probing the interaction between dark matter (DM) and the Standard Model (SM) particles. It provides an alternative to the direct and indirect DM searches in probing the DM parameter space. The advantage of the LHC search is its independency of the cosmological assumptions.
Because DM is invisible to the LHC detectors, its LHC signature involves a visible object $X$ recoiling against the invisible DM. 
The object $X$ depends on the details of the DM interaction with the SM sector. 
One of the earlier considered scenario is the interaction of DM with the light quarks in the effective field theory (EFT) framework. 
In this scenario, the DM recoils against the initial state radiation (ISR) coming from the quarks. The strongest bounds comes from the ISR jet, hence monojet signature~\cite{Goodman:2010yf, Bai:2010hh, Goodman:2010ku, Fox:2012ee, Barger:2012pf, CMS:2017tbk, ATLAS:2017dnw, Khachatryan:2016reg}, while the monophoton signatures also provide a competitive bound~\cite{Fox:2011pm, Zhou:2013fla, Sirunyan:2017ewk, Aaboud:2017dor}.
It was soon realized that the EFT framework is not applicable at the LHC energy scale where the typical momentum transfer in DM-SM interaction is comparable to the suppression scale of the EFT.  
In order to avoid working with a specific model, simplified models~\cite{Frandsen:2012rk, Busoni:2013lha, DiFranzo:2013vra, Buchmueller:2013dya, Busoni:2014sya, Papucci:2014iwa, Buchmueller:2014yoa, Abdallah:2014hon, Buckley:2014fba, Jacques:2015zha, Primulando:2015lfa, Buchmueller:2015eea,  Abdallah:2015ter, DeSimone:2016fbz, Englert:2016joy, Bell:2016uhg, Kahlhoefer:2015bea, Ellis:2017tkh} are introduced to correctly taking into account the kinematics of DM interactions. 
The simplified models introduce mediators between DM and the SM sector. 
This further enriches collider signatures of the dark sector. 
In some regions of parameter space, the search for the mediators provides a better bound than the search for DM alone.

Besides DM coupling to light quarks there are other possibilities of the DM-SM couplings. The couplings of the DM to the third generation quarks are studied in the mono-$b$, the $b\bar b$ + MET and the $t \bar t $ + MET channels~\cite{Lin:2013sca, Sirunyan:2017xgm, Aad:2014vea, Chen:2016ylf}. The DM-Higgs coupling is probed by a monoHiggs signature~\cite{Petrov:2013nia, Aaboud:2017yqz, Sirunyan:2017hnk}. Mono-$Z$ and mono-$W$ are used to probe the DM-weak gauge boson coupling~\cite{Bai:2012xg, Cotta:2012nj,Sirunyan:2017onm, Khachatryan:2016mdm, Bell:2015rdw, Haisch:2016usn, Chen:2013gya}.

In all the cases mentioned above, the signal are generally of order $\alpha_{dark}\alpha_{s(ew)}$ where the irreducible backgrounds coming from the neutrino is of order $\alpha_{ew}\alpha_{s(ew)}$. The choice between $\alpha_s$ or $\alpha_{ew}$ depends on whether the final states involve quark/gluon or lepton/weak gauge boson/Higgs. From this one can see that the LHC can probe the DM-SM coupling $\alpha_{dark}$ up to the $\alpha_{ew}$. The same estimation cannot be said for the case of dark matter recoiling against a neutrino. For example in the monolepton case, the background is of order $\alpha_{ew}$ while the signal is of order $\alpha_{dark}\alpha_{ew}$. From this estimation, it seems that in this case the LHC bounds on $\alpha_{dark}$ are not perturbative. However in this work we will demonstrate that the kinematic distributions between the signal and the background are different. Therefore we can impose kinematics cuts to reduce the background significantly. By recasting the data from the 13 TeV mono-lepton search~\cite{Aaboud:2017efa}, we show that the bounds on the DM-neutrino is still in the perturbative range for a large extend of DM parameter space.

Model independent interaction of dark matter and neutrino can also be constrained from cosmological searches. The interaction of DM and the neutrino might results in a suppression of the primordial density perturbations which can be seen in the CMB or the matter power spectra~\cite{Boehm:2000gq, Mangano:2006mp, Serra:2009uu, Wilkinson:2014ksa, Escudero:2015yka,Shoemaker:2013tda}. 
By analyzing the matter power spectrum, Ref.~\cite{Wilkinson:2014ksa} obtained the upper bounds on the present day value of DM-neutrino scattering cross-section of $\sigma_{\text{DM}+\nu \rightarrow \text{DM}+\nu} \lesssim  10^{-33} \frac{m_{\text{DM}}}{\text{GeV}} \text{ cm}^2$ if the cross section is constant and $\sigma_{\text{DM}+\nu \rightarrow \text{DM}+\nu} \lesssim 10^{-45} \frac{m_{\text{DM}}}{\text{GeV}} \text{ cm}^2$, if the cross section depends on the square of the temperature.
Neutrino telescopes also provide some bounds on the interaction from the DM annihilation into neutrinos. The current bounds are $\langle \sigma_{\text{DM}\text{DM}\rightarrow \nu\nu} v\rangle \lesssim \mathcal O \left(10^{-24} - 10^{-23}\right) \text{cm}^3/\text{s}$ for DM mass 0.01 to 100 GeV. We found that the LHC run-2 bounds are competitive with these cosmological bounds. 

This paper is organized as follows. In Sec~\ref{sec:framework} we set up our framework and notations for studying the DM-neutrino interactions. We also derive phenomenological consequences of the DM-neutrino interaction in our setup. In Sec.~\ref{sec:ColliderBounds} we discuss LHC signatures of our DM-neutrino interactions and the current LHC bounds. We next turn to discuss the bounds on DM-neutrino interactions from neutrino telescopes in Sec.~\ref{sec:NeutrinoBounds}. We briefly discuss the mediators and possible UV completions in Sec.~\ref{sec:Mediator}. We then conclude in Sec.~\ref{sec:conclusion}. 

\section{Minimal dark matter-neutrino interaction}
\label{sec:framework}

When considering the DM-neutrino interaction, it is expected that the EFT framework will fail to properly describe the kinematics at the LHC environment. Therefore we will use the simplified model approach in this paper. 
Two types of simplified model are widely discussed in the literature: the s-channel mediator and the t-channel mediator. In the s-channel mediator scenario, the LHC only put constraints on the neutrino-mediator coupling. The DM-mediator coupling is unconstrained, hence an assumption on its value have to be made to compare the LHC and the cosmological bounds. In the t-channel mediator framework, the DM-neutrino-mediator interaction is parametrized by a single coupling. Here no additional assumptions are needed to compare the cosmological and the LHC constraints. We will use this t-channel mediator assumption for the remainder of this paper.

We make an assumption that DM, $\chi$, a mediator, $\phi$ (or $\psi$ for a fermionic mediator), and the SM fields are the only light degrees of freedom. 
The rest of new physics particles are heavy and decoupled. 
In this simplified scenario, the renormalizable interaction between DM and a neutrino is given by  
\begin{equation} \label{eq:lag}
\mathcal L_{int} = \left\{\begin{aligned}&y\, \bar\chi P_L \nu \phi + \text{h.c.},\\
	&y\,\bar\psi P_L \nu\chi + \text{h.c.},\end{aligned}\right.
	\quad \begin{aligned} \text{fermionic DM},&\\ \text{scalar DM},&\end{aligned}
\end{equation}
where $y$ is the DM-neutrino coupling constant. For simplicity we assume that $y$ is real. We also assume that the coupling is flavor democratic.
For $\chi$ to be stable, we must have $m_\chi \le m_{\phi(\psi)}$.\footnote{Note that $\chi$ stands for both the scalar and fermionic dark matter.}
In principle, DM can be its own antiparticle if $\chi$ is a Majorana fermion or a real scalar. However, for definiteness, in this work we'll assume that DM is not its own antiparticle.
The cosmological constraints for the scalar case has been considered in Ref. \cite{hm:2017mio}.

In this setup, we'll allow for the possibility that $\chi$ and $\phi(\psi)$ are not their own antiparticles. 
Generically, both $\chi$ and $\phi(\psi)$ could be part of an electroweak multiplet.
However, in this work we'll leave their electroweak charge assignment unspecified except that they're neutral under $U(1)_{EM}$. 
We'll briefly revisit the topic of their possible charge assignments and UV completions in Sec~\ref{sec:Mediator}.

In the rest of this section, we explore the phenomenological consequences of the DM-neutrino interaction in  Eq.~\eqref{eq:lag}.
\subsection{Neutrinos signal from DM annihilation}
\label{sec:nusignal}
\begin{figure}
       \centering
        \subfloat[Fermionic DM]{\includegraphics[width= 0.3\textwidth]{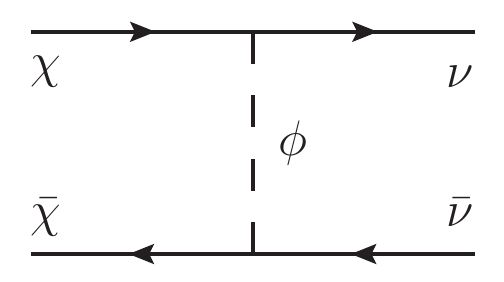}}
        \hspace{1cm}
        \subfloat[Scalar DM]{\includegraphics[width= 0.3\textwidth]{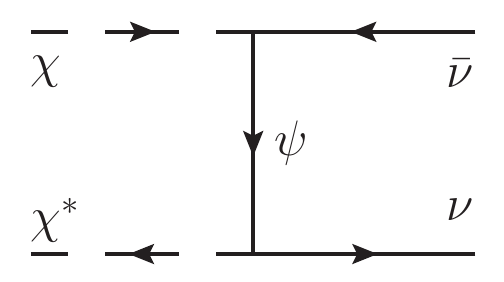}}
        \caption{The DM self annihilation diagram for the case of fermionic DM and scalar DM.}
         \label{fig:xsec}
\end{figure}

In our simplified scenario of DM-neutrino interaction given in Eq.~\eqref{eq:lag}, a pair of DM particle can annihilate into a pair of neutrinos, see Fig.~\ref{fig:xsec}.
The DM self annihilation cross-section  reads
\begin{equation}
\begin{aligned}
	\sigma v_{rel} = \left\{
	\begin{aligned}
		&\frac{9 y^4}{32\pi}\frac{m_\chi^2}{(m_\chi^2+m_\phi^2)^2} + \mathcal{O}(v_{rel}^2),\\
		&\frac{3 y^4}{16\pi}\frac{m_\chi^2}{(m_\chi^2+m_\psi^2)^2}v_{rel}^2 + \mathcal{O}(v_{rel}^4),
	\end{aligned}
	\qquad
	\begin{aligned}
		\text{fermionic DM}&,\phantom{\frac{a^2_2}{b^2_2}}\\
		\text{scalar DM}&,\phantom{\frac{a^2_2}{b^2_2}}
	\end{aligned}
	\right.
\end{aligned}
\label{eq:xsec}
\end{equation}
where $v_{rel}$ is the DM relative velocity. 
Note in the above expression we have summed over the 3 neutrino species.

The neutrinos produced by DM annihilation will each carries energy $m_\chi$. They can be looked for at neutrino telescopes  such as IceCube, KamLAND and Super-Kamiokande.
The flux of a particular neutrino (or an antineutrino) flavor from the DM self-annihilation observed on Earth can be written as~\cite{Yuksel:2007ac}
\begin{equation}
	\frac{d\Phi_{\nu_f}}{dE_\nu d\Omega} = \frac{d\Phi_{\bar\nu_f}}{dE_{\bar\nu} d\Omega} = \frac{\langle\sigma v_{rel} \rangle}4 \mathcal{J}_{avg} \frac{R_{sc}\rho_{sc}^2}{4\pi m_{\chi}^2}\frac{1}{3}\delta(E_{\nu(\bar\nu)}-m_\chi),
	\label{eq:diffflux}
\end{equation}
where $f$ is a lepton family index, $R_{sc} = 8.5$ kpc is the Solar radius~\cite{2016ApJS..227....5D} and $\rho_{sc}$ is the DM density at the Solar radius. 
The factor of 1/3 comes from the assumption that each neutrino flavor is equally likely to be produced. Note in the case where DM is its own antiparticle, the factor 1/4 in the above equation becomes 1/2.
The neutrinos from the annihilation point will oscillate in flight on the way to the detector. 
With the present neutrino oscillation parameters, we expect the flavor ratio of the neutrino at the detection point to be $\nu_e:\nu_\mu:\nu_\tau\simeq1:1:1$~\cite{PalomaresRuiz:2007eu}. 
The factor $\mathcal{J}_{avg}$ represent the average line-of-sight integral over the whole sky
\begin{equation}
	\mathcal{J}_{avg}  =  \frac{1}{2R_{sc}\rho_{sc}^2}\int_{-1}^1 d\left(\cos\theta\right)\int_0^{l_{max}} dl \rho^2 \left(\sqrt{R_{sc}^2 - 2 l R_{sc}\cos\theta + l^2}\right),
\end{equation}
where $\theta$ is the angle subtended from the DM annihilation point to the Galactic Center, $\rho(r)$ is the DM density profile as a function of distance from the Galactic Center and $l$ is the distance from the annihilation point to the Earth. The limit of integration depends on the size of the DM halo and is given by $l_{max} = \sqrt{R_{halo}^2-R_{sc}^2\,\sin^2\theta} + R_{sc}\cos\theta$. In our calculation, we take $R_{halo} = 100$ kpc.

\begin{table}
\begin{centering}
\begin{tabular}{| l | l | l |}
\hline
Parameter & NFW &Burkert\\
\hline
$(\alpha,\beta,\gamma,\delta)$ & $(1,3,1,0)$ & $(2,3,1,1)$\\[0.1cm]
$r_s$ [kpc] & $16.1^{+17.0}_{-7.8}$ & $9.26^{+5.6}_{-4.2}$\\[0.1cm]
$\rho_{sc}$ [GeV/$\text{cm}^3$] & $0.471^{+0.048}_{-0.061}$ & $0.487^{+0.075}_{-0.088}$\\[0.1cm]
\hline
\end{tabular}
\caption{Parameters for the NFW and the Burkert DM density profiles~\cite{Nesti:2013uwa}}
\label{tab:profile}
\end{centering}
\end{table} 

The dark matter profile can be parametrized as
\begin{equation}
	\rho(r) = \frac{\rho_0}{\left(\delta+\frac{r}{r_s}\right)^\gamma\left[1+\left(\frac{r}{r_s}\right)^\alpha\right]^{(\beta-\gamma)/\alpha}},
\end{equation}
where $\rho_0$ is the density normalization deduced from the DM density at the Solar radius ($\rho_{sc}$), $r_s$ is the scale factor and $\alpha,\beta,\gamma,\delta$ are the shape parameters.
In this work, we choose the Navaro, Frank and White (NFW)~\cite{Navarro:1995iw} and the Burkert~\cite{Burkert:1995yz} profiles as representative DM profiles. 
The NFW profile represents a cuspy profile while the Burkert profile represents profile with a core.  
The value of the parameters for both the NFW and the Burkert profiles are given in Tab.~\ref{tab:profile}.
Using the central values for the parameter, the $\mathcal{J}_{avg}$ values are determined to be $\mathcal{J}_{avg} = 3.34$ for the NFW profile and  $\mathcal{J}_{avg} = 1.60$ for the Burkert profiles. These values agree with Ref.~\cite{Khatun:2017adx}. 

The bounds from neutrino telescopes on DM-neutrino interaction will be discussed in detailed in Sec.~\ref{sec:NeutrinoBounds}.

\subsection{DM-neutrino scattering}
\label{sec:chinuscattering}
\begin{figure}
       \centering
        \subfloat[Fermionic DM]{\includegraphics[width= 0.23\textwidth]{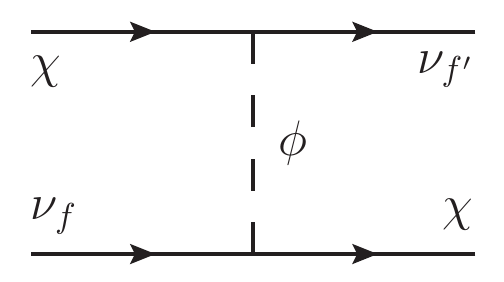}
        \includegraphics[width= 0.2\textwidth]{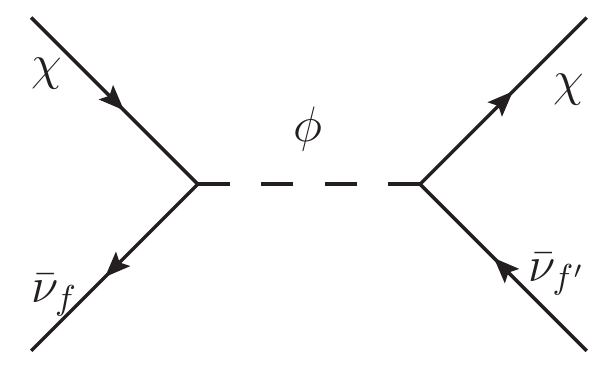}}
        \hspace{.1cm}
        \subfloat[Scalar DM]{\includegraphics[width= 0.20\textwidth]{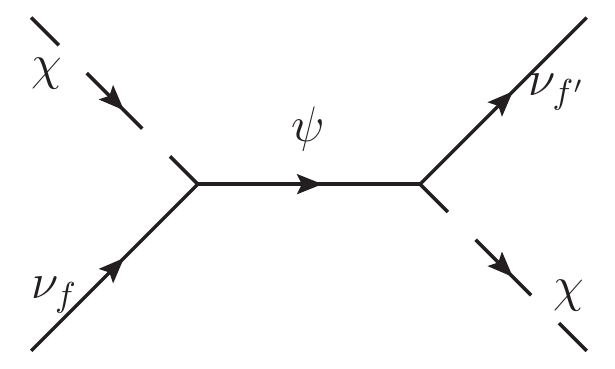}
         \includegraphics[width= 0.23\textwidth]{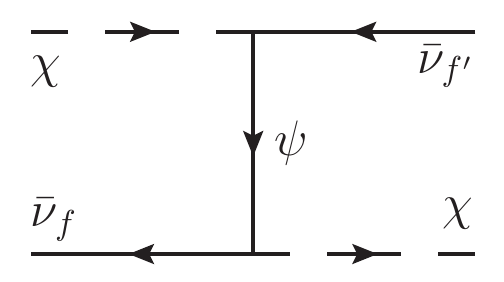}}
        \caption{The diagrams contribute to $\chi-\nu$ and $\chi-\bar\nu$ scattering for the fermionic DM case (a) and the scalar DM case (b) }
        \label{fig:chinu}
\end{figure}
\begin{figure}
       \centering
        \includegraphics[width= 0.5\textwidth]{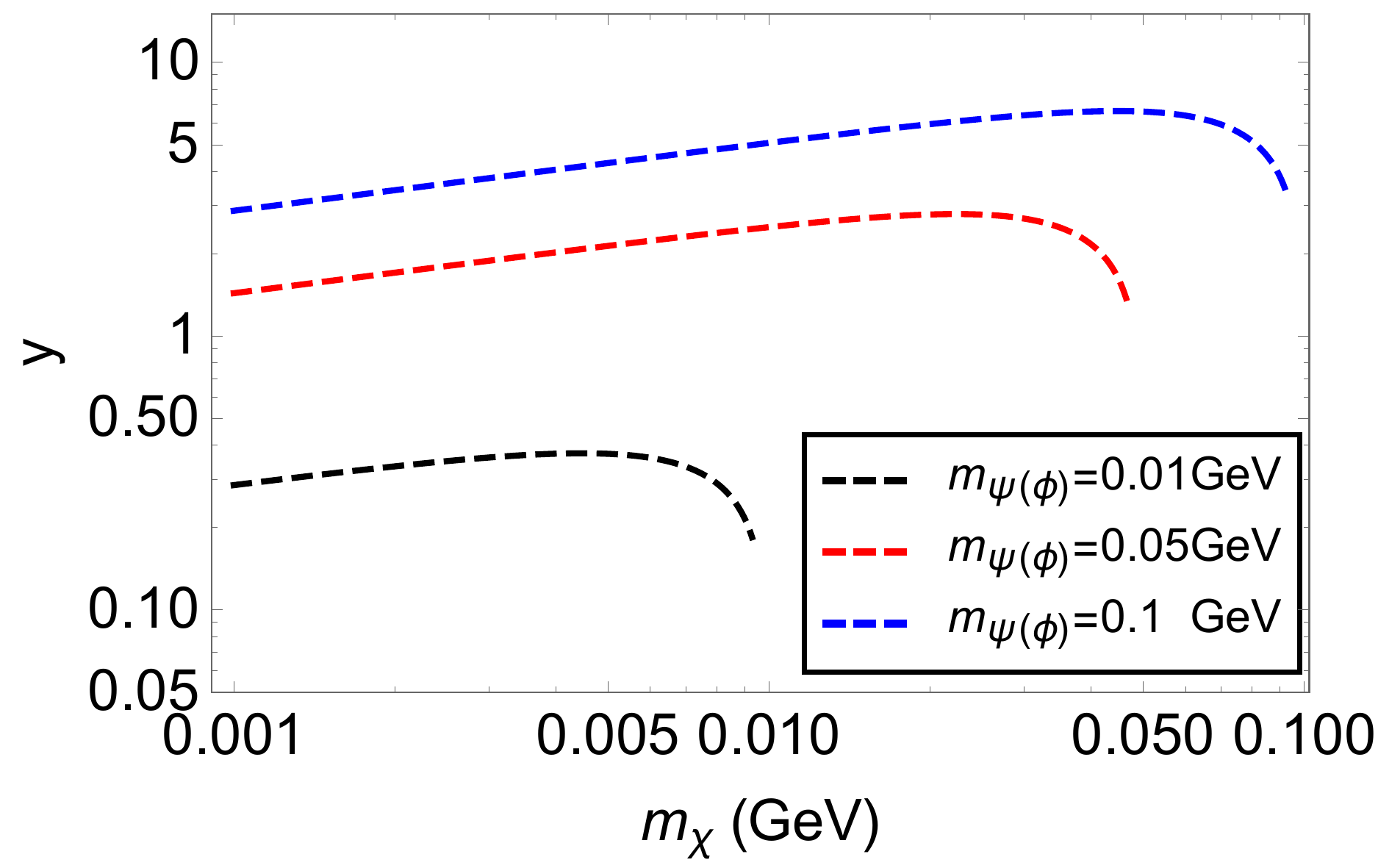}
        \caption{The cosmological bound on the DM-neutrino scattering cross-section as a function of the DM mass for various value of messenger masses. The dip at the end of each line corresponds to the increase in DM-neutrino scattering as $m_\chi\to m_\phi$.}
        \label{fig:cosmoboundy}
\end{figure}

The DM-neutrino interaction in Eq.~\eqref{eq:lag} induce DM-neutrino scattering via Feynman diagram shown in Fig.~\ref{fig:chinu}. For the fermionic DM case, the DM-neutrino cross-section is given by
\begin{equation}
	\sigma(\chi\nu\to\chi\nu) = \sigma(\chi\bar\nu\to\chi\bar\nu) = \frac{3y^4}{16\pi}\frac{E_\nu^2}{(m_\phi^2-m_\chi^2)^2},
\end{equation} 
where we have sum over the 3 neutrino species in the final state. The cross-section for the scalar DM case is the same up to a trivial replacement $m_\phi\to m_\psi$. This scattering process leaves an imprint on CMB and matter power spectra which affects the large-scale structure of the universe. The most stringent cosmological constraint on DM-neutrino interaction comes from the matter power spectrum, $\sigma(\chi\nu\to\chi\nu)<10^{-45}\frac{m_\chi}{\text{GeV}}\text{ cm}^2$~\cite{Wilkinson:2014ksa}. 
It translates to the bound on the DM-neutrino coupling as a function of the DM and the messenger mass. Fig.~\ref{fig:cosmoboundy} shows such a bound for the benchmark case of messenger mass of 0.01, 0.05 and 0.1 GeV respectively. Since the $\chi$-$\nu$ scattering cross-section scales roughly as $1/m_{\phi(\psi)}^4$, the bound on the coupling becomes weak at large messenger mass as can be seen from the plot. Similarly, one can translate the bound on $\chi$-$\nu$ cross-section into the bound on DM annihilation cross-section, see Fig.~\ref{fig:nfwbound} and~\ref{fig:burkertbound}. 

\subsection{New decay channels for $Z$ and $W$}
\begin{figure}
       \centering
        \subfloat[$Z\to\nu\chi\phi$]{\includegraphics[width= 0.3\textwidth]{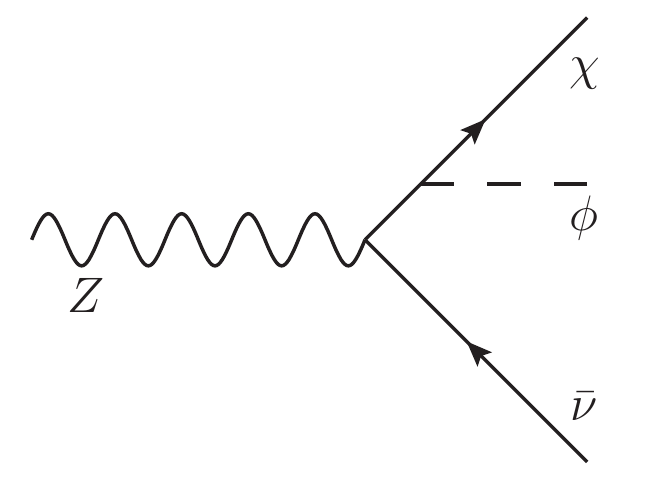}
        \includegraphics[width= 0.3\textwidth]{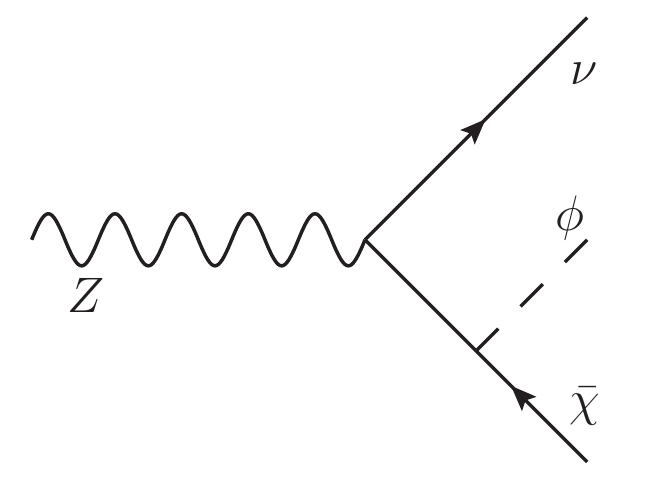}}
        \hspace{.7cm}
        \subfloat[$W^+\to\ell^+\chi\phi$]{\includegraphics[width= 0.3\textwidth]{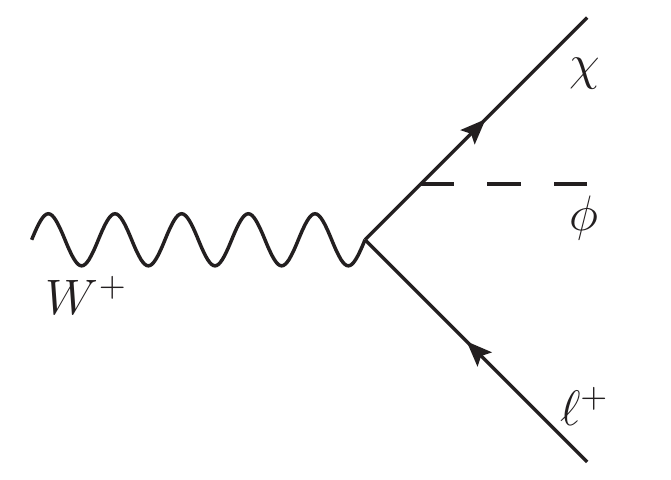}}
        \caption{The diagrams contribute to (a) the invisible decay of the $Z$ and (b) the mono-lepton decay of the $W^+$ for the case of fermionic DM. Similar diagrams for the case of scalar DM can be obtained by making the replacement $\chi\to\psi$ and $\phi\to\chi$ in the above diagrams.}
        \label{fig:3bodydecay}
\end{figure}

The DM-neutrino interaction in Eq.~\eqref{eq:lag} could lead to a new decay channel for both the $Z$ and the $W$ boson if it is kinematically open. In the case of fermionic DM, the new decay channels are $Z\to\nu\chi\phi$ and $W\to\ell\chi\phi$.
The Feynman diagrams for such a process are shown in Fig.~\ref{fig:3bodydecay}.
Since both the $\chi$ and $\phi$ are neutral, they would appear invisible to the detector. Hence, the $Z\to\nu\chi\phi$ decay contributes to the invisible decay width of the $Z$,\footnote{In the case $m_\chi+m_\phi$ just heavier than $m_Z$, the $Z$ can still decay invisibly into $\nu\chi\phi^\ast\to2\nu2\chi$.  We have found that this channel does not give stronger constraints compared with other search channels for our benchmark points.}
while the $W\to\ell\chi\phi$ contributes to the mono-lepton decay width of the $W$.
The differential partial decay width for $Z\to\nu\chi\phi$ is given by
\begin{equation}
\begin{aligned}
	\frac{d\Gamma(Z\to\nu\chi\phi)}{dm_{\nu\chi}^2dm_{\chi\phi}^2} &= \frac{1}{(2\pi)^3}\frac{y^2}{16v_{ew}^2m_{Z}^3m_{\chi\phi}^4}
	\big(2m_{Z}^4(m_{\chi\phi}^2+m_\chi^2-m_\phi^2) \\
	&\hspace{3cm}+ m_{\chi\phi}^4(m_{\nu\chi}^2-m_\chi^2)  - 2m_{Z}^2m_{\chi\phi}^2(m_{\nu\chi}^2+m_{\chi\phi}^2-m_\phi^2)\big),
\end{aligned}
\end{equation}
where $m^2_{\nu\chi} = (p_\nu+p_\chi)^2$, $m^2_{\chi\phi} = (p_\chi+p_\phi)^2$, $v_{ew}$ is the electroweak vev and $m_Z$ is the $Z$ boson mass. 
The limits on the kinematic variables are
\begin{align}
	m_\chi^2&\le m^2_{\nu\chi} \le (m_{Z}-m_\phi)^2,\\
	(m_{\chi\phi}^2)_{\text{min,max}} &= \left(p_\chi^{(0)}+p_\phi^{(0)}\right)^2 - \left(\sqrt{\left(p_\chi^{(0)}\right)^2-m_\chi^2}\pm\sqrt{\left(p_\phi^{(0)}\right)^2-m_\phi^2}\right)^2,
\end{align}
where 
\begin{equation}
	p_\chi^{(0)} = \frac{m_{\nu\chi}^2+m_\chi^2}{2m_{\nu\chi}},\qquad
	p_\phi^{(0)} = \frac{m_{Z}^2-m_{\nu\chi}^2-m_\phi^2}{2m_{\nu\chi}}.
\end{equation}
The current invisible branching ratio of the $Z$ boson is $(20.00\pm0.06)\%$~\cite{Olive:2016xmw}.

The differential partial decay rate for $W\to\ell\chi\phi$ is given by
\begin{align}
	\frac{d\Gamma(W\to\ell\chi\phi)}{dm_{\ell\chi}^2dm_{\chi\phi}^2} &= \frac{1}{(2\pi)^3}\frac{y^2}{32v_{ew}^2 m_W^3 m_{\chi\phi}^4}\left[2m_{W}^4(m_{\chi\phi}^2+m_\chi^2-m_\phi^2) + m_{\chi\phi}^4(m_{\ell\chi}^2-m_\chi^2-2m_{W}^2) \right.\nn\\
	&\hspace{2.5cm}\left.  + m_{W}^2m_{\chi\phi}^2(2m_\phi^2+m_\ell^2-2m_{\ell\chi}^2)\right],
\end{align}
where $m_W$ is the $W$ boson mass and 
\begin{align}
	(m_\chi+m_\ell)^2 &\le m_{\ell\chi}^2 \le (m_{W}-m_\phi)^2,\\
	(m_{\chi\phi}^2)_{\text{min,max}} &= \left(p_\chi^{(0)}+p_\phi^{(0)}\right)^2 - \left(\sqrt{\left(p_\chi^{(0)}\right)^2-m_\chi^2}\pm\sqrt{\left(p_\phi^{(0)}\right)^2-m_\phi^2}\right)^2.
\end{align}
The variables $p_\chi^{(0)}$ and $p_\phi^{(0)}$, in this case, are given by
\begin{equation}
	p_\chi^{(0)} = \frac{m_{\ell\chi}^2+m_\chi^2-m_\ell^2}{2m_{\ell\chi}},\qquad
	p_\phi^{(0)} = \frac{m_{W}^2-m_{\ell\chi}^2-m_\phi^2}{2m_{\ell\chi}}.
\end{equation}
The current mono-lepton branching ratio of the $W$ boson is $(32.72\pm0.30)\%$~\cite{Olive:2016xmw}.

For the case of scalar DM, the expression for the differential partial decay width of the $Z$ and the $W$ boson can be obtained by an obvious substitution $\phi\to\psi$.

Moreover, for a hadron collider such as the LHC, the $Z\to\nu\chi\phi(\psi)$ and $W\to\ell\chi\phi(\psi)$ decay channels inevitably lead to the mono-jet signal and the mono-lepton signal respectively. 
These collider signatures are actively being investigated at the LHC. 
We'll discuss the collider bound in details in the next section.

\section{LHC Bounds}
\label{sec:ColliderBounds}

At the LHC, the neutrinos are produced copiously from the $W$ and $Z$ decays. If the neutrinos interact with the dark sector according to Eq.~\eqref{eq:lag}, the DM and the mediator will also be produced at the LHC, as shown in Fig.~\ref{fig:3bodydecay}. While both the DM and the mediator are invisible at the LHC, this interaction could change the kinematics of $W$ and $Z$ productions.  In the rest of this section we will assume the fermionic DM case for definiteness. The scalar DM case can be obtained with a trivial replacement $\phi\to\psi$ for the mediator.

\begin{figure}
       \centering
        \includegraphics[width= 0.7\textwidth]{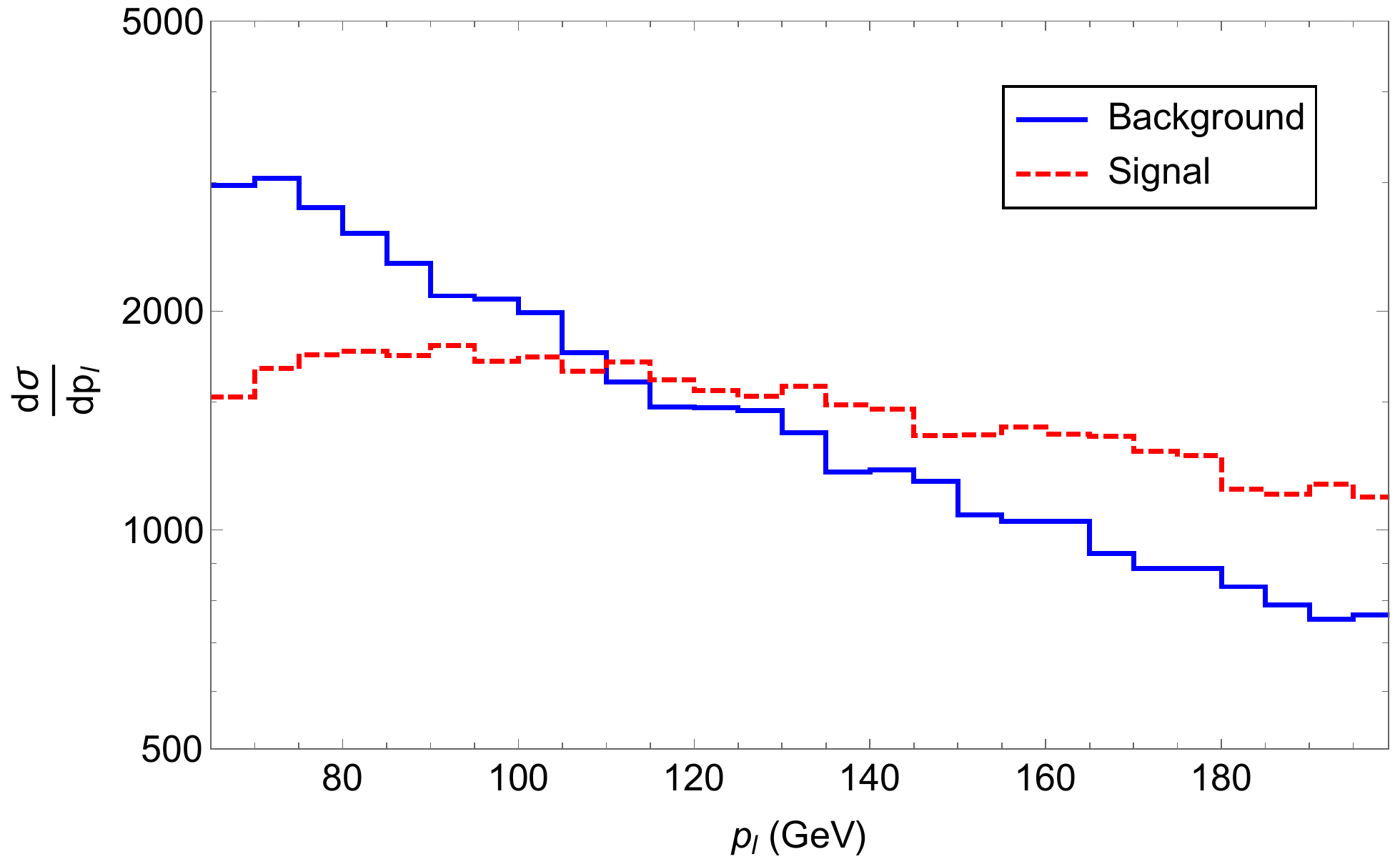}
        \caption{The lepton momentum distribution of the background ($\ell\nu$) and signal ($\ell\phi\chi$) at the partonic level obtained from the Madgraph simulation.  For the signal, we use $m_\chi = 30$ GeV and $m_\phi = 100$ GeV. The normalization of the cross section is arbitrary.}
         \label{fig:emomentum}
\end{figure}

The LHC signature for the dark sector production via $W$-boson is a lepton and missing energy from $\phi$ and $\chi$, dubbed monolepton. The cuts imposed by the ATLAS or CMS force the $W$ to be off-shell. Hence the distribution of the lepton momentum can be approximated as
\begin{equation}
\frac{d\sigma_{pp\rightarrow \ell \phi\chi}}{dp_\ell} \propto \frac{p_\ell}{s \, m_{\phi\chi}^2}, 
\end{equation}
where $p_\ell$ is the lepton momentum, $s$ is the center-of-mass energy and $m_{\phi\chi}$ is the invariant mass of the $\phi$ and $\chi$. The background for this process is the production of lepton and neutrino. In this case, the lepton momentum distribution is  
\begin{equation}
\frac{d\sigma_{pp\rightarrow \ell\nu}}{dp_\ell} \propto \frac{\delta\left( p_\ell - \sqrt{s}/2 \right)}{s}.
\end{equation}
Taking into account the parton distribution function, the lepton momentum distributions are shown in Fig.~\ref{fig:emomentum}. From the plot, one can see that the lepton momentum for the signal tends to be harder than the background. 

If $\phi$ and $\chi$ are produced via $Z$-boson, they need an object to recoil against to show a LHC signature. The most promising search channel for this scenario is the monojet channel. However we found that this channel is not as sensitive as the monolepton channel. The monojet channel is a dirtier channel for the LHC environment. Hence commonly ATLAS and CMS puts strong cuts on the jet $p_T$ hence the signal acceptance is reduced. For this paper, we only consider monolepton channel. Combining the monolepton channel with monojet channel does not change the results appreciably. 
  
We base our analysis on the 13 TeV, 36.1 fb$^{-1}$ ATLAS monolepton search~\cite{Aaboud:2017efa}. In the electron channel, the ATLAS analysis requires exactly one electron with $p_T >$ 65 GeV and $|\eta| < 1.37$ or $1.52 <|\eta|<2.47$. The requirement for the muon channel is exactly one muon with $p_T > 55$ GeV with $|\eta| < 2.5$. The missing energy has to be greater than 65(55) GeV in the electron(muon) channel. In both channels, the events are vetoed if there is an additional lepton with $p_T > 20$ GeV. The signal and the backgrounds are discriminated by the transverse mass distributions. The transverse mass variable is defined as $M_T = \sqrt{2p_T E_T^{miss}\left(1-\cos\phi_{\ell\nu}\right)}$, where $p_T$ is the transverse momentum of the lepton, $E_T^{miss}$ is the missing transverse energy, and $\phi_{\ell\nu}$ is the angle between the lepton and the missing energy in the transverse plane. 

\begin{figure}
       \centering
        \subfloat[Electron channel]{\includegraphics[width= 0.5\textwidth]{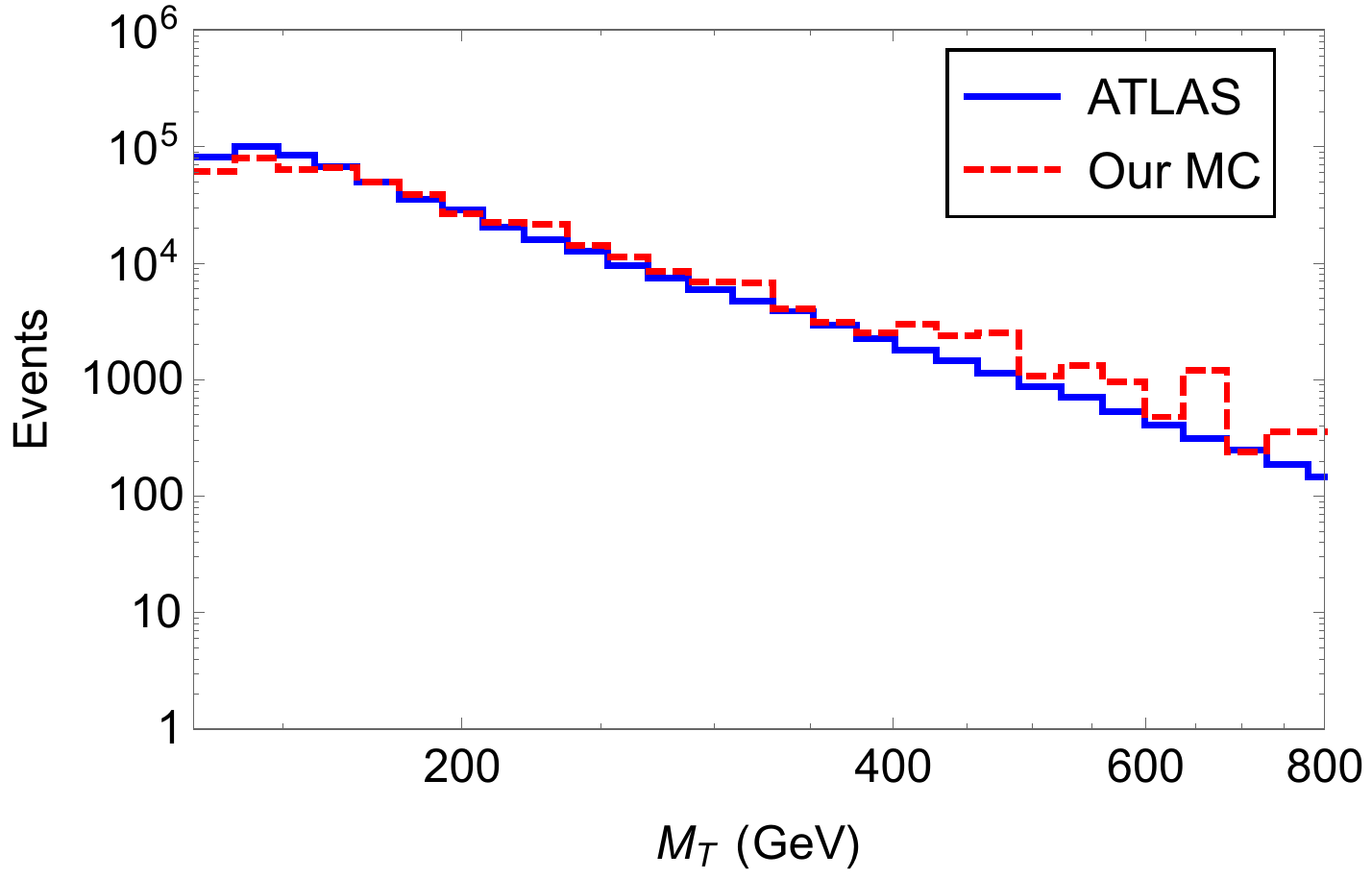}}
        \subfloat[Muon channel]{\includegraphics[width= 0.5\textwidth]{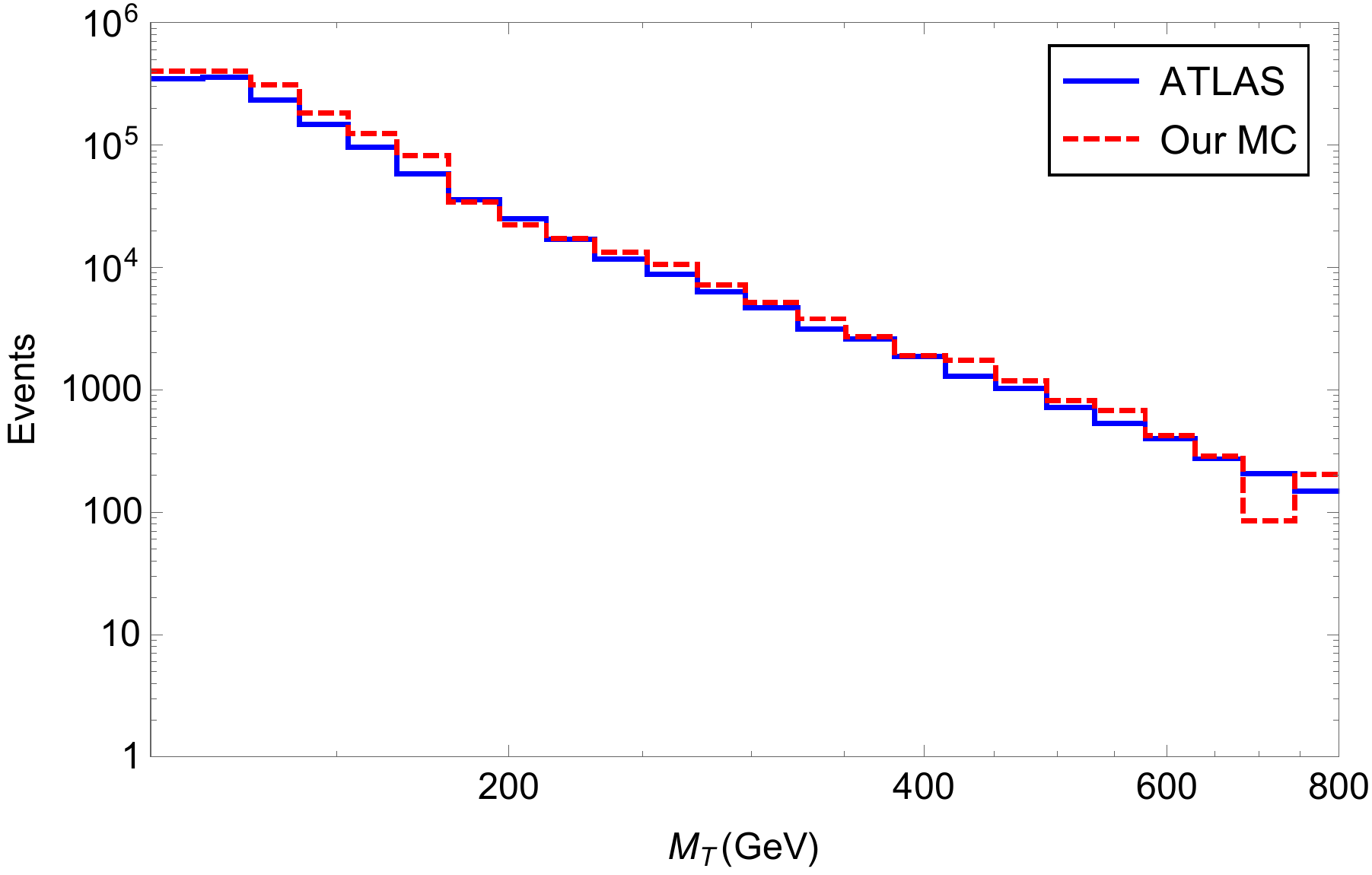}}
        \caption{The comparison between the ATLAS and our background simulation for the monolepton analysis ($W \rightarrow \ell \nu$).}
         \label{fig:wlnu}
\end{figure}

\begin{figure}
       \centering
        \subfloat[Electron channel]{\includegraphics[width= 0.5\textwidth]{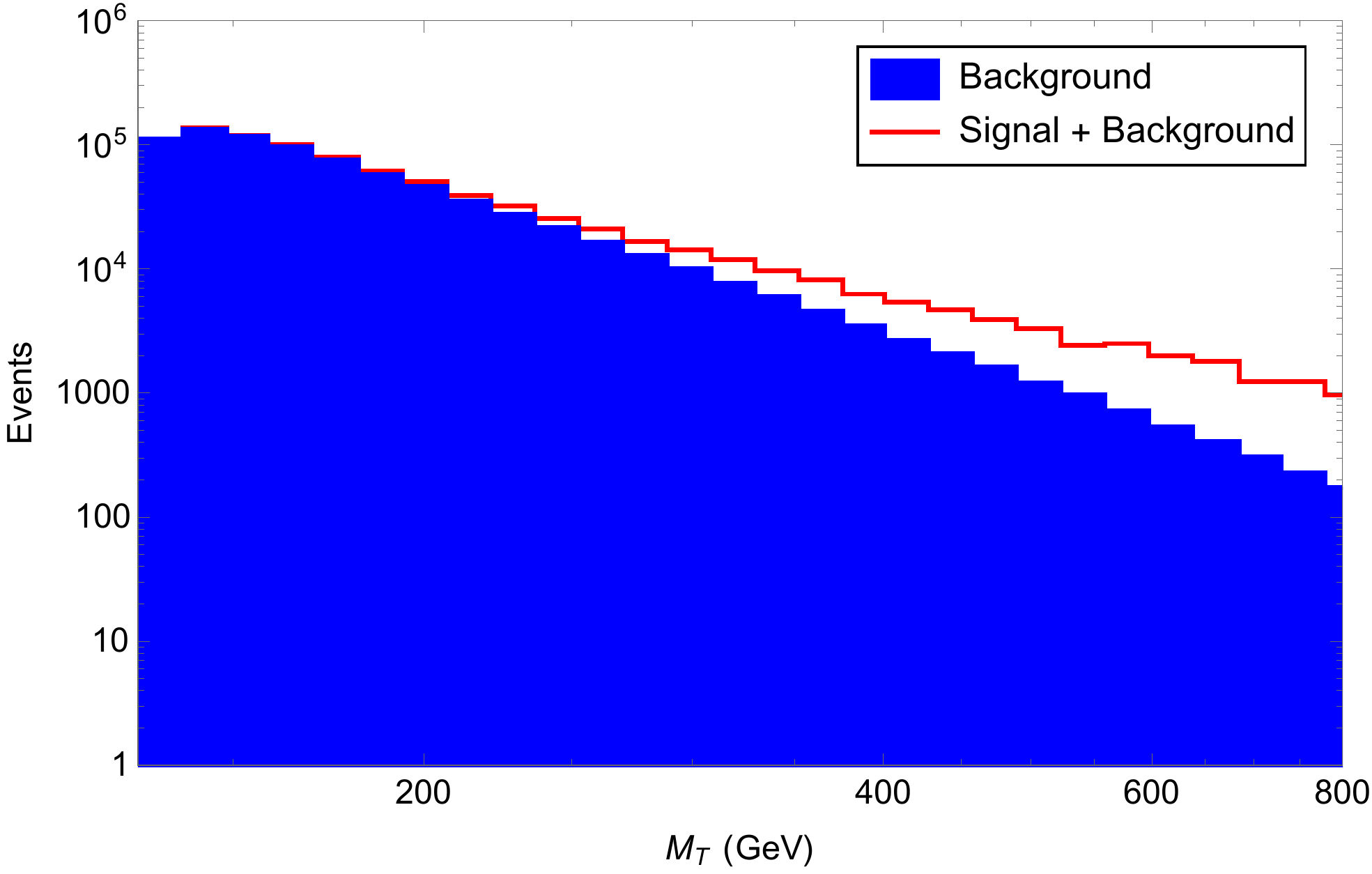}}
        \subfloat[Muon channel]{\includegraphics[width= 0.5\textwidth]{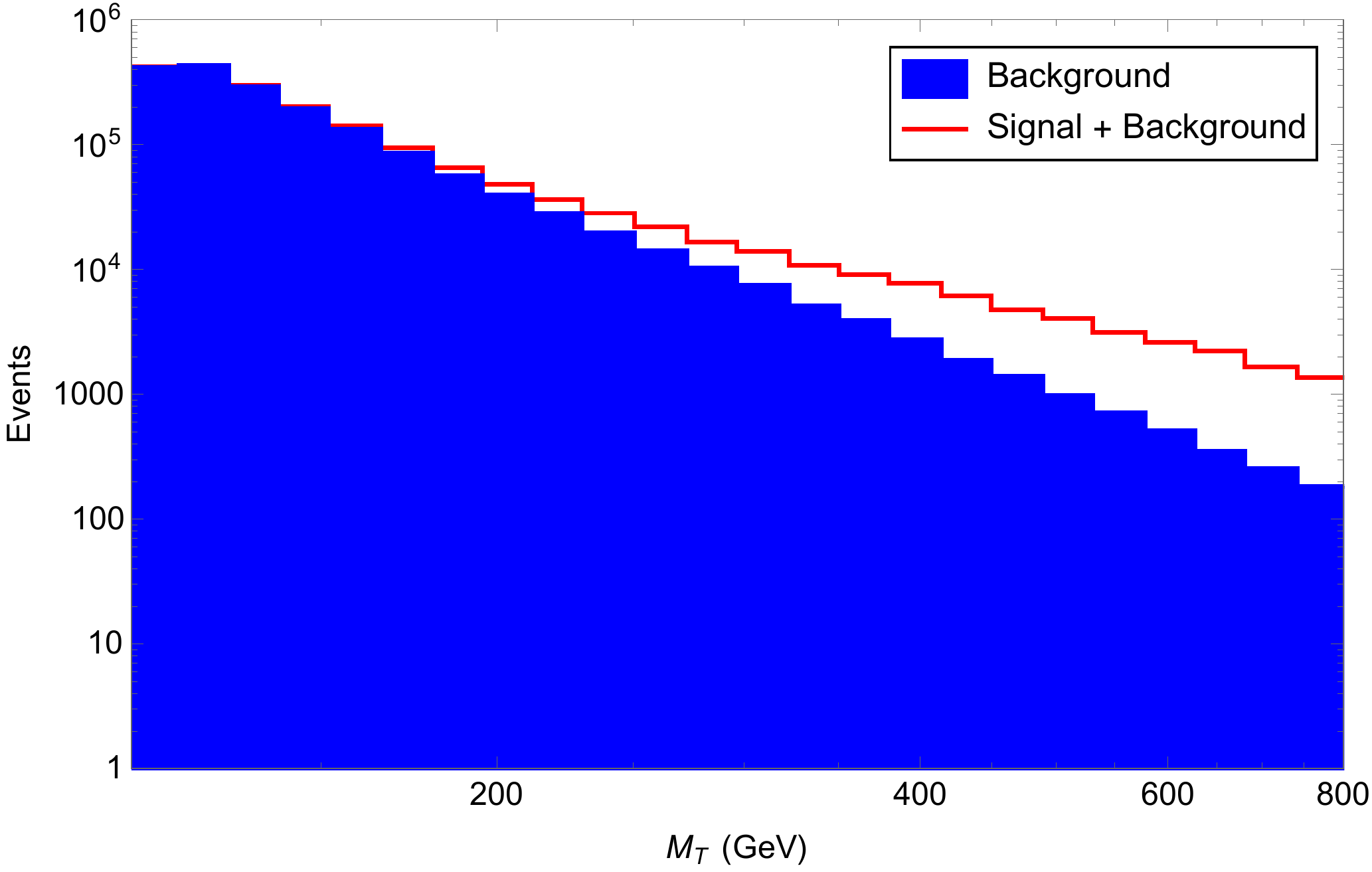}}
        \caption{The comparison between the background and the signal+background distributions. For the signal, we take $m_\chi = 30$ GeV and $m_\phi = 100$ GeV with an arbitrary normalization.}
         \label{fig:sigback}
\end{figure}

Both of the background and the signal events are simulated using \texttt{MadGraph 5}~\cite{Alwall:2014hca}, followed by matching, parton shower and hadronization using \texttt{Pythia 6}~\cite{Sjostrand:2006za}. The detector simulation and pileup are simulated using \texttt{Delphes 3}~\cite{deFavereau:2013fsa}. The average number of pileup in the simulation is 23. The signal model is generated using \texttt{Feynrules 2.0}~\cite{Alloul:2013bka}. The comparison for the ATLAS and our simulation for the background process $W \rightarrow \ell \nu$ are shown in Fig.~\ref{fig:wlnu}. The signal transverse mass distribution compared with the total background is given in Fig.~\ref{fig:sigback}.

\begin{figure}
       \centering
        \subfloat[Fermion DM]{\includegraphics[width= 0.5\textwidth]{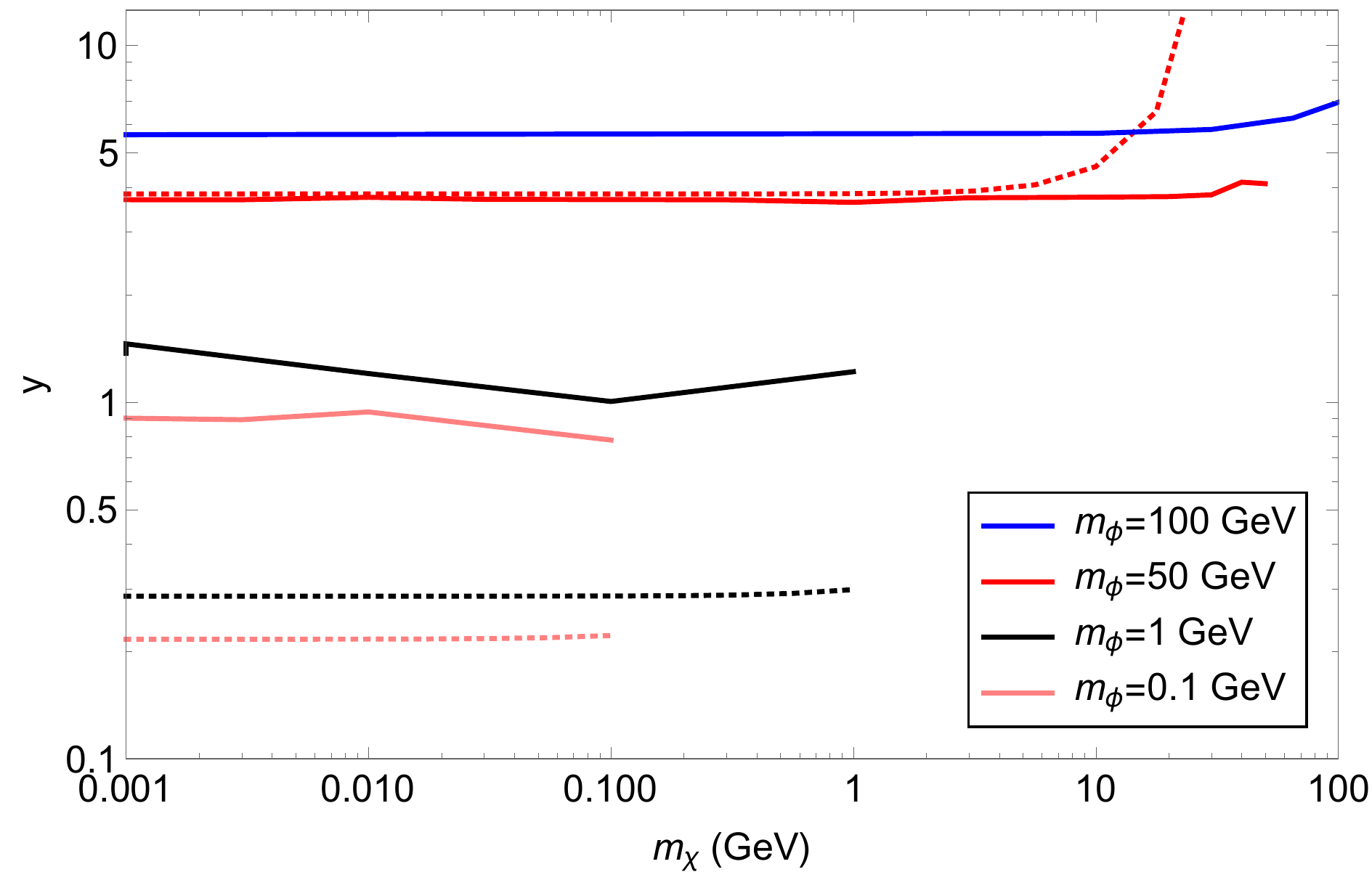}}
        \subfloat[Scalar DM]{\includegraphics[width= 0.5\textwidth]{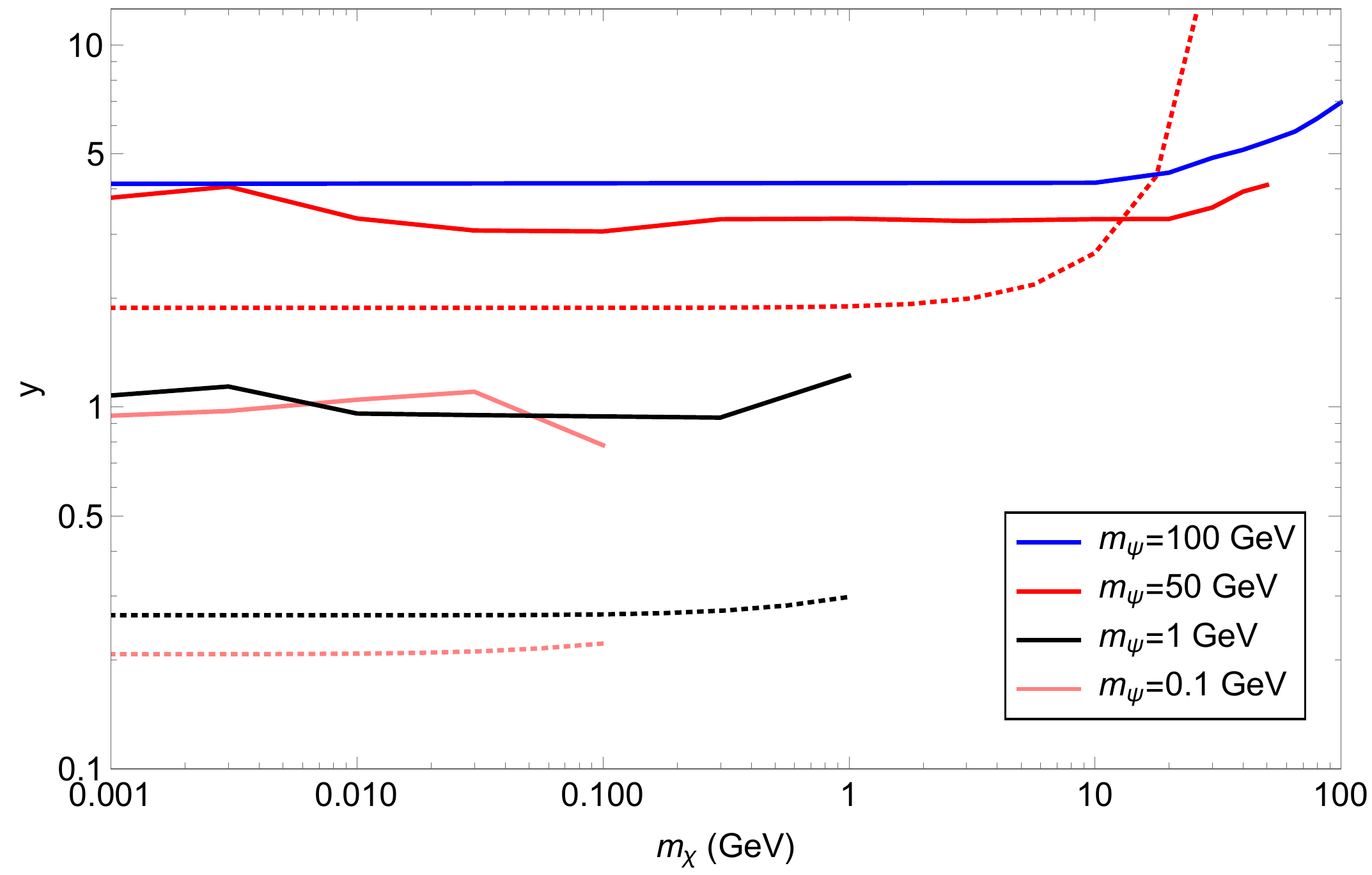}}
        \caption{The 90\% CL bounds on the DM-neutrino coupling $y$ from the LHC mono-lepton search (solid line) and the invisible $Z$ decay width (dotted line).  The collider bounds are subjected to $y\leq 4\pi$ and $m_\chi \leq m_{\phi(\psi)}$ constraints.}
         \label{fig:ybound}
\end{figure}

ATLAS collaboration bin their results according to the transverse mass. There are seven bins, starting from $m_T =$130(110) GeV for electron(muon) channel to 7 TeV. To get the bounds, we compared the signal and the background in each bin. The likelihood of observing the signal in a bin is given by 
\begin{equation}
\mathcal L_i\left(n_i|s_i,b_i\right)=\int_0^\infty \frac{\left(\xi\left(s_i+b_i\right)\right)^{n_i} e^{-\xi\left(s_i+b_i\right)}}{n_i!} P_i(\xi) d\xi,
\label{eq:likelihood}
\end{equation}
where $b_i$ is the number of predicted background in the bin, $n_i$ is the number of observed events in the bin, $s_i$ is the number of predicted signal in the bin which is a function of the coupling $y$. The function $P_i(\xi)$ is introduced to incorporate the uncertainty in each bin and take a log-normal form~\cite{Balazs:2017moi}
\begin{equation}
P_i\left(\xi\right) = \frac{1}{\sqrt{2\pi} \sigma_i}\frac{1}{\xi}\exp\left(-\frac{1}{2}\left(\frac{\ln \xi}{\sigma_i}\right)^2\right),
\end{equation}
where $\sigma_i$ is the systematics error for the corresponding bin. The chi-squared value is given by
\begin{equation}
\chi^2 = 2\sum_{i} \left(\ln\mathcal L_i\left(n_i|s_i,b_i\right) - \ln\mathcal L_i\left(n_i|s_i = 0,b_i\right)\right),
\label{eq:chi2}
\end{equation}
We combine the bounds from both of the electron  and muon channels. 
The 90\% confidence level (CL) bounds are shown in Fig.~\ref{fig:ybound}. 
There, we consider the bounds for both the fermionic and the scalar dark matter cases with various dark matter masses. For the benchmark points, the mediator mass are chosen to be 0.1 GeV, 1 GeV, 50 GeV and 100 GeV. We also show bounds from the invisible $Z$ decay width. From the plot, one can see that the LHC bounds become more competitive for a higher mediator mass, where the $Z$ decay is starting to become kinematically non-favorable. 
These bounds can be compared with the indirect detection bounds discussed in the next section.

\section{Indirect Detection Bounds}
\label{sec:NeutrinoBounds}
The DM-neutrino interaction  gives rise to a pair annihilation of DM particles into a pair of neutrinos. 
These neutrinos, in principle, can be observed at the neutrino telescopes on Earth. 
For example, the IceCube experiment searches for neutrinos from DM annihilation in the Milky Way for DM mass range between 30 GeV and 10 TeV~\cite{Aartsen:2016pfc}, with the updated IceCube analysis covers DM mass range between 10 GeV and 1 TeV~\cite{Aartsen:2017ulx}. These two IceCube studies place an upper limit on the DM self annihilation cross-section, $\langle\sigma v\rangle\lesssim \mathcal{O}(10^{-23})\text{ cm}^3/$s. Similar analysis has been carried out by the Super-Kamiokande (SK) experiment for DM mass ranging from 1 GeV to 10 TeV~\cite{Frankiewicz:2015zma}, resulting in the upper bound $\langle\sigma v\rangle\lesssim \mathcal{O}(10^{-24}-10^{-23})\text{ cm}^3/$s. Additionally, the KamLAND experiment has searched for neutrinos from DM annihilation for a low DM mass region from 8.3 MeV to 31.8 MeV and places an upper limit of $\langle\sigma v\rangle\lesssim \mathcal{O}(10^{-24})\text{ cm}^3/$s~\cite{Collaboration:2011jza}.

In additional to the dedicated neutrinos from DM annihilation searches, one can also derive the upper bound on DM annihilation cross-section from other existing neutrino searches. In particular, we will focus on reinterpreting the SK supernova relic neutrino search~\cite{Bays:2011si} and the SK atmospheric neutrino measurement~\cite{Richard:2015aua} in the context of DM annihilation.

Additionally, neutrino telescopes data could be used to bound the DM-neutrino cross-section, which translates to the constraint on DM-neutrino coupling. Ref.~\cite{Arguelles:2017atb} obtained an upper bound on DM-neutrino coupling from analysing the IceCube high-energy neutrino data set~\cite{Aartsen:2015zva}. Such a bound is relevant for high DM mass and/or high messenger mass and can be complimentary to the bounds obtained in this work.
\subsection{Bounds from Super-Kamiokande Supernova Relic Neutrino Searches}
\label{sk-snr}
A supernova explosion throughout the history of the universe release most of its energy in the form of neutrinos. 
These neutrinos, referred to as the supernova relic neutrino (SRN), should remain in existence today.
The SRN signal can be detected at the SK detector via the inverse beta decay process, where $\bar\nu_e$ scatters off a free proton inside the SK detector into a neutron and a positron.   
These SRN data can be reinterpreted to derive a constraint on DM annihilation cross-section for DM mass in the range of 10 $\sim$ 130 MeV. 
Ref.~\cite{PalomaresRuiz:2007eu} carried out such an analysis on the first 1497 days of SK data (SK-I)~\cite{Malek:2002ns}. 
In this subsection, we give the updated analysis by including data from the second (SK-II) and the third (SK-III) data taking period of the SK detector~\cite{Bays:2011si}. 
Our analysis differs from that of Ref.~\cite{PalomaresRuiz:2007eu} in the treatment of the SK energy resolution and signal efficiency which we describe below. Also in the case of a fermionic DM, we take DM to be a Dirac particle instead of a Majorana fermion.  As a result, our bounds come out to be similar to the one obtained in Ref.~\cite{PalomaresRuiz:2007eu} even though we use roughly twice as much data.

The SK SRN searches detect $e^+$ from the inverse beta decay process in the energy range 18-82 MeV. 
The detected $e^+$ is then binned into 16 four-MeV energy bins. 
To take into account the energy resolution of the SK detector, we introduce the energy resolution function
\begin{equation}
	R\left(E,E'\right) = \frac{1}{\sqrt{2\pi} \sigma} \exp\left(-\frac{\left(E-E'\right)^2}{2\sigma^2}\right),
\end{equation} 
where $E$ is the actual $e^+$ energy, $E'$ is the detected energy and $\sigma$ is the SK detector resolution which is given in~\cite{Hosaka:2005um}\footnote{Ref.~\cite{Hosaka:2005um} gives the energy resolution up to 20 MeV. For a higher value of $e^+$ energy, we conservatively estimate the energy resolution to be the same as the resolution at 20 MeV, $\sim$ 11.4\%. }.

The expected number of $e^+$ from DM self-annihilation in the energy bin $i$ is given by 
\begin{equation}
	N_i = 4\pi \mathcal N_{SK} \mathcal T \int dE_{\bar\nu_e}  \frac{d\Phi_{\bar\nu_e}}{dE_{\bar\nu_e}d\Omega} \int_{E_{i-1}}^{E_{i}} dE'_{e} \int_{E_{min}}^{E_{max}} dE_{e}\,R\left(E^{\phantom{\prime}}_e,E_e'\right)  \mathcal{E}\, \frac{d\sigma_{\bar\nu_e p}}{dE_{e}},
\end{equation}
where $\mathcal N_{SK}=1.5\times10^{33}$ is the number of proton in the SK fiducial volume, $\mathcal T$ is the total running time, $E_0=18$ MeV, $E_{i}-E_{i-1}=4$ MeV, $E_{min(max)}$ is the minimum (maximum) actual $e^+$ energy, $\mathcal E$ is the detector efficiency which is a function of positron energy, and $\sigma_{\bar\nu_e p}$ is the inverse beta decay cross-section. In our computation, we take $\sigma_{\bar\nu_e p}$ from Ref.~\cite{Strumia:2003zx}. 
 
In order to derive the upper limit on DM self annihilation cross-section from the SK data, we introduce the likelihood function for each energy bin  as in Eq.~\eqref{eq:likelihood}. 
Then we construct the $\chi^2$ function, as in Eq.~\eqref{eq:chi2}.
Note in this case the sum is over all energy bins and over the three SK data taking periods. 
Finally, the 90\% CL upper limit on DM self-annihilation cross-section is determined from $\Delta\chi^2=2.71$.

\subsection{Bounds from Super-Kamiokande Atmospheric Neutrino Measurements}
\label{sk-atm}
The SK experiment measures the atmospheric neutrino fluxes in the energy range from 100 MeV to 100 GeV. 
The SK detector is sensitive to both the electron- and the muon-neutrinos. 
However, the detector cannot distinguish neutrinos from antineutrinos, thus SK reports the fluxes of $\nu_e+\bar\nu_e$ and $\nu_\mu+\bar\nu_\mu$ separately.
The latest SK atmospheric neutrino measurements, which include SK I-IV data set, is reported in Ref.~\cite{Richard:2015aua}. 
In these measurements, the observed neutrino fluxes as well as their uncertainties are reported. 

The SK atmospheric neutrino measurements can be reinterpret to derive a bound on DM annihilation cross-section for DM mass between 100 MeV and 100 GeV. We use the method of least square to determine the bound. The $\chi^2$ function is defined as
\begin{equation}
	\chi^2 = \sum_i\frac{\left( \Phi^{\nu_e,i}_{atm} + \Phi^{\nu_e,i}_{DM} - \Phi^{\nu_e,i}_{obs}\right)^2}{\left(\sigma^{\nu_e,i}\right)^2}+\frac{\left( \Phi^{\nu_\mu,i}_{atm} + \Phi^{\nu_\mu,i}_{DM} - \Phi^{\nu_\mu,i}_{obs}\right)^2}{\left(\sigma^{\nu_\mu,i}\right)^2},
\end{equation}
where $\Phi^{\nu_{e(\mu)},i}_{atm}$ is the predicted atmospheric electron (muon) neutrino flux for bin $i$, $\Phi^{\nu_{e(\mu)},i}_{obs}$ is the observed atmospheric electron (muon) neutrino flux at bin $i$ and $\sigma^{\nu_{e(\mu)},i}$ is the uncertainty associated with bin $i$. 
This includes both the uncertainty in the measurement (statistical and systematics) as well as the uncertainty in the predicted atmospheric neutrino flux. 
The differential neutrinos flux per solid unit angle from DM annihilation in bin $i$, $d\Phi^{\nu_{e({\mu})},i}_{DM}/d\Omega$, according to Eq.~\eqref{eq:diffflux} can be is written as 
\begin{equation}
	\frac{d\Phi^{\nu_{e({\mu})},i}_{DM}}{d\Omega} = \int_{E_{min}}^{E_{max}} dE_\nu \frac{\langle\sigma v \rangle}{4}  \mathcal{J}_{avg} \frac{R_{sc}\rho_{sc}^2}{4\pi m_\chi^2}\frac{2}{3}\delta(E_\nu-m_\chi),
\end{equation}
where $E_{min}$ and $E_{max}$ is the minimum and maximum energy of each bin respectively and the factor of 2 come from combining both the neutrinos and the antineutrinos. Note that in the case where DM is its own antiparticle, the factor of 1/4 in the cross-section becomes 1/2. We take the value of the predicted atmospheric neutrino flux from the HKKM simulation~\cite{Honda:2011nf}. We estimate the uncertainty in the predicted atmospheric neutrino flux by comparing it against the Bartol~\cite{Barr:2004br} and FLUKA~\cite{Battistoni:2002ew} simulations.

\subsection{Comparison of Bounds}
\begin{figure}
       \centering
                \subfloat[Fermion DM]{\includegraphics[width= 0.49\textwidth]{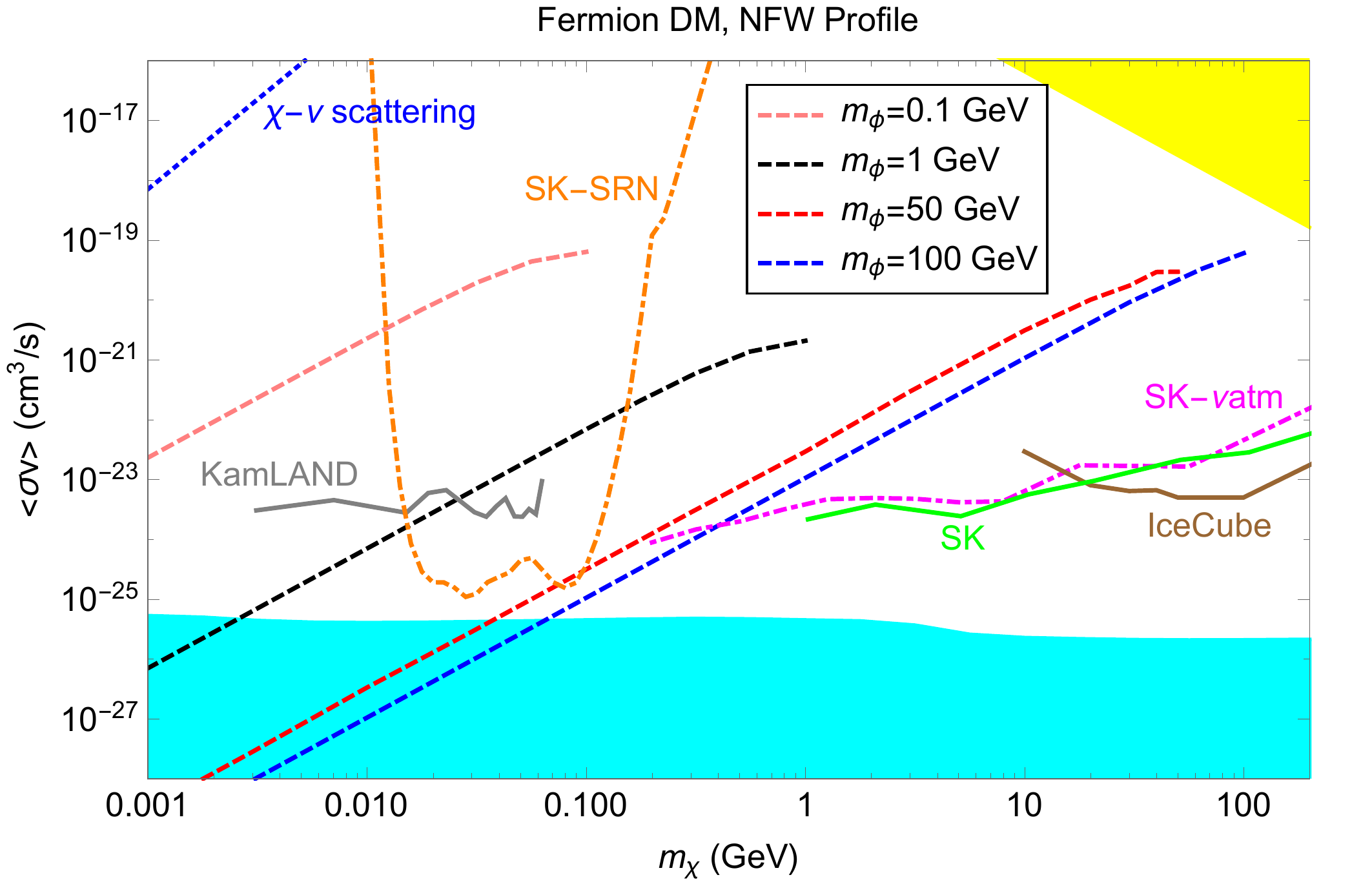}\label{fig:nfwfermion}} 
        \subfloat[Scalar DM with $v_{rel} = 10^{-3}$]{\includegraphics[width= 0.49\textwidth]{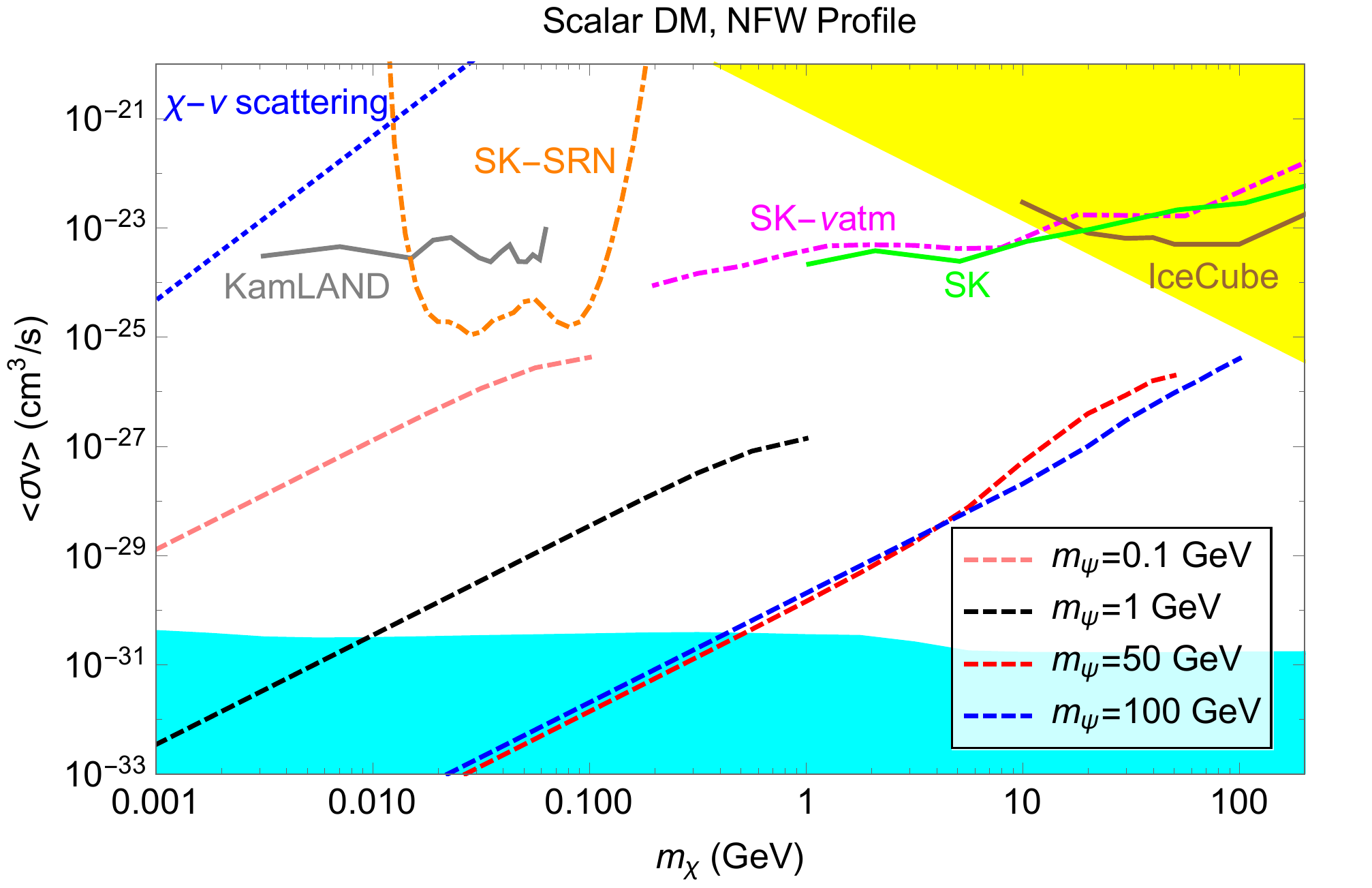}\label{fig:nfwscalar}} 
        \caption{The 90\% CL bounds from various neutrino telescope observations compared with the collider bounds. The gray line, labelled KamLAND, is the (rescaled) official KamLAND bound~\cite{Collaboration:2011jza}. The green line, labelled SK, is the official SK bound~\cite{Frankiewicz:2015zma}. The brown line, labelled IceCube, is the official IceCube bound~\cite{Aartsen:2017ulx}. The orange, dot-dashed line, labelled SK-SRN, is the bound obtained from recasting the SK SRN searches discussed in Sec.~\ref{sk-snr}. The magenta, dot-dashed line, labelled SK-$\nu$atm, is the bound obtained from recasting the SK atmospheric neutrino measurements discussed in Sec.~\ref{sk-atm}. The shaded yellow region is excluded by perturbativity constraint on the DM-neutrino coupling, $y$, and the constraint $m_\chi\le m_{\phi(\psi)}$. The shaded cyan region is the region that gives $\Omega_\chi h^2 > 0.1186$ simulated using Micromegas 5.0 \cite{Belanger:2018ccd}. The blue dotted lines are the cosmological bounds from the DM-neutrino scattering discussed in Sec~\ref{sec:chinuscattering}. For the collider bounds (dashed lines), we only show strongest bound between the LHC and the invisible $Z$ decay width bounds. We subject the collider bounds to $y\leq 4\pi$ and $m_\chi \leq m_{\phi(\psi)}$. In these plots, the dark matter profile is taken to be the NFW profile.}
         \label{fig:nfwbound}
\end{figure}

\begin{figure}
       \centering
        \subfloat[Fermion DM]{\includegraphics[width= 0.49\textwidth]{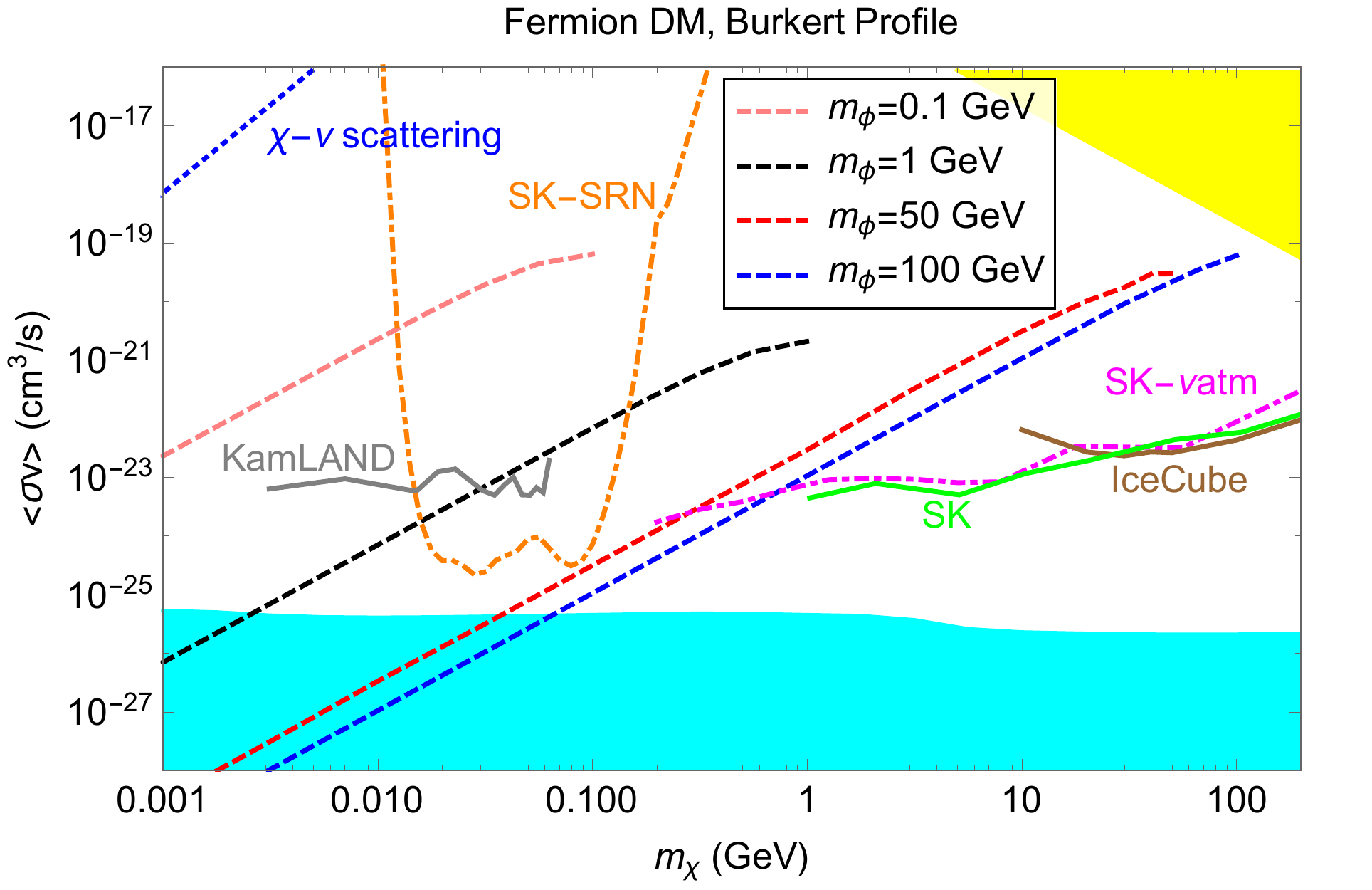}\label{fig:burkertfermion}} 
        \subfloat[Scalar DM with $v_{rel} = 10^{-3}$]{\includegraphics[width= 0.49\textwidth]{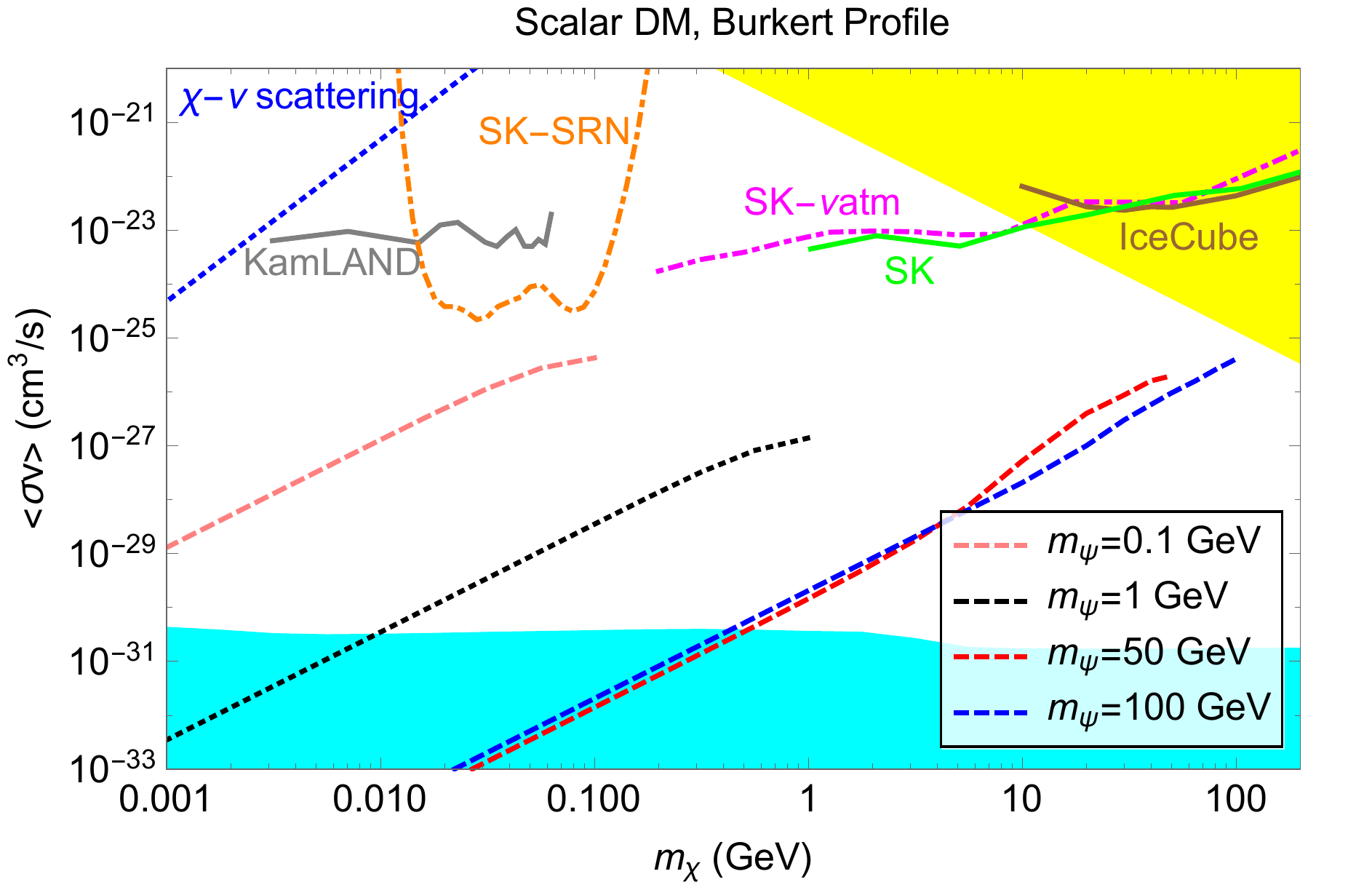}\label{fig:burkertscalar}} 
        \caption{The 90\% CL bounds from various neutrino telescope observations compared with the collider bounds. The gray line, labelled KamLAND, is the (rescaled) official KamLAND bound~\cite{Collaboration:2011jza}. The green line, labelled SK, is the (rescaled) official SK bound~\cite{Frankiewicz:2015zma}. The brown line, labelled IceCube, is the official IceCube bound~\cite{Aartsen:2017ulx}. The orange, dot-dashed line, labelled SK-SRN, is the bound obtained from recasting the SK SRN searches discussed in Sec.~\ref{sk-snr}. The magenta, dot-dashed line, labelled SK-$\nu$atm, is the bound obtained from recasting the SK atmospheric neutrino measurements discussed in Sec.~\ref{sk-atm}.  The shaded yellow region is excluded by perturbativity constraint on the DM-neutrino coupling, $y$, and the constraint $m_\chi\le m_{\phi(\psi)}$. The shaded cyan region is the region that gives $\Omega_\chi h^2 > 0.1186$ simulated using Micromegas 5.0 \cite{Belanger:2018ccd}. The blue dotted lines are the cosmological bounds from the DM-neutrino scattering discussed in Sec~\ref{sec:chinuscattering}. For the collider bounds (dashed lines), we only show strongest bound between the LHC and the invisible $Z$ decay width bounds. We subject the collider bounds to $y\leq 4\pi$ and $m_\chi \leq m_{\phi(\psi)}$.  In these plots, the dark matter profile is taken to be the Burkert profile.}
         \label{fig:burkertbound}
\end{figure}

\begin{figure}
       \centering
        \subfloat[$y = 1$]{\includegraphics[width= 0.49\textwidth]{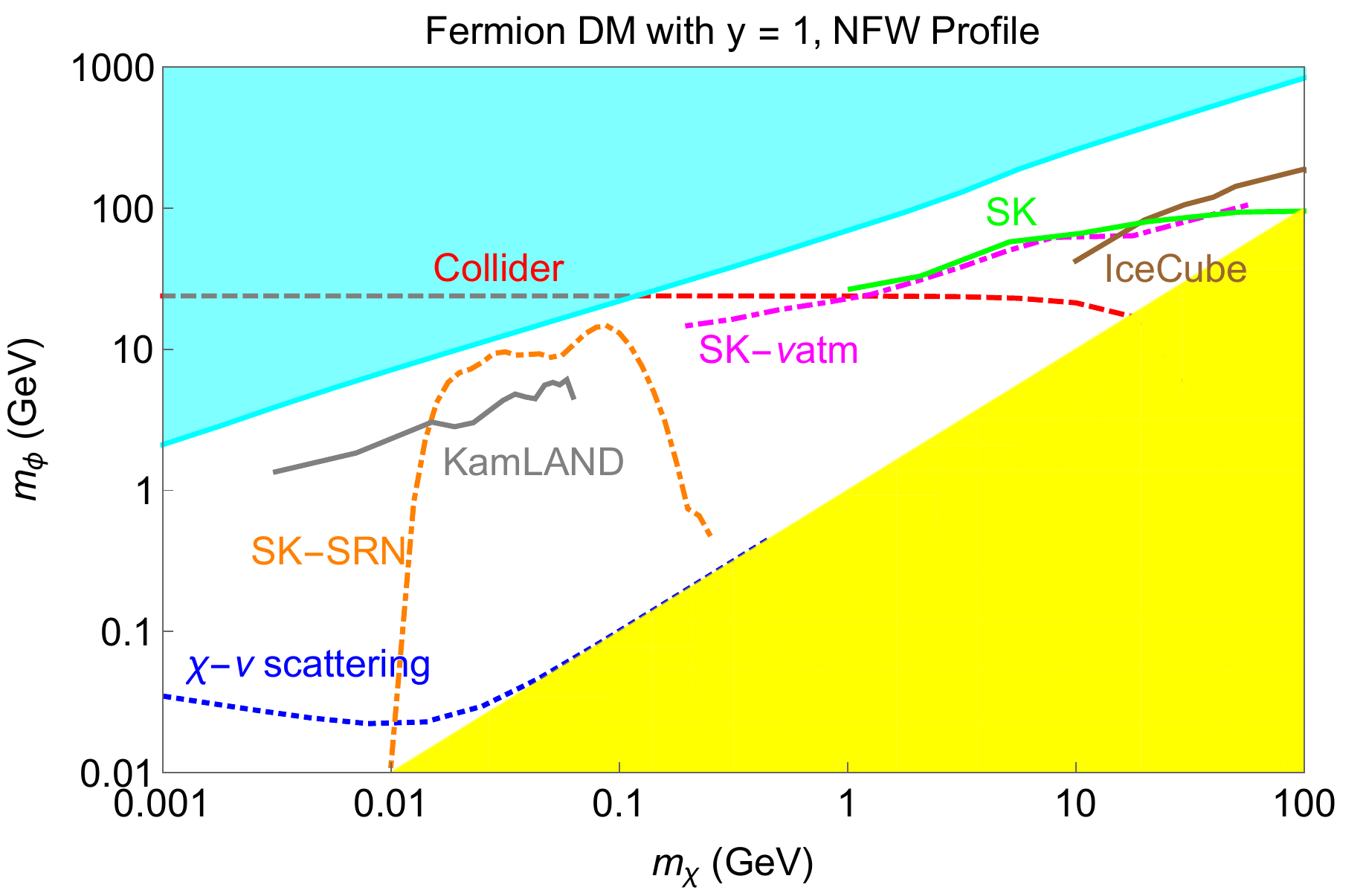}\label{fig:chiphi1}} 
        \subfloat[$y=4$]{\includegraphics[width= 0.49\textwidth]{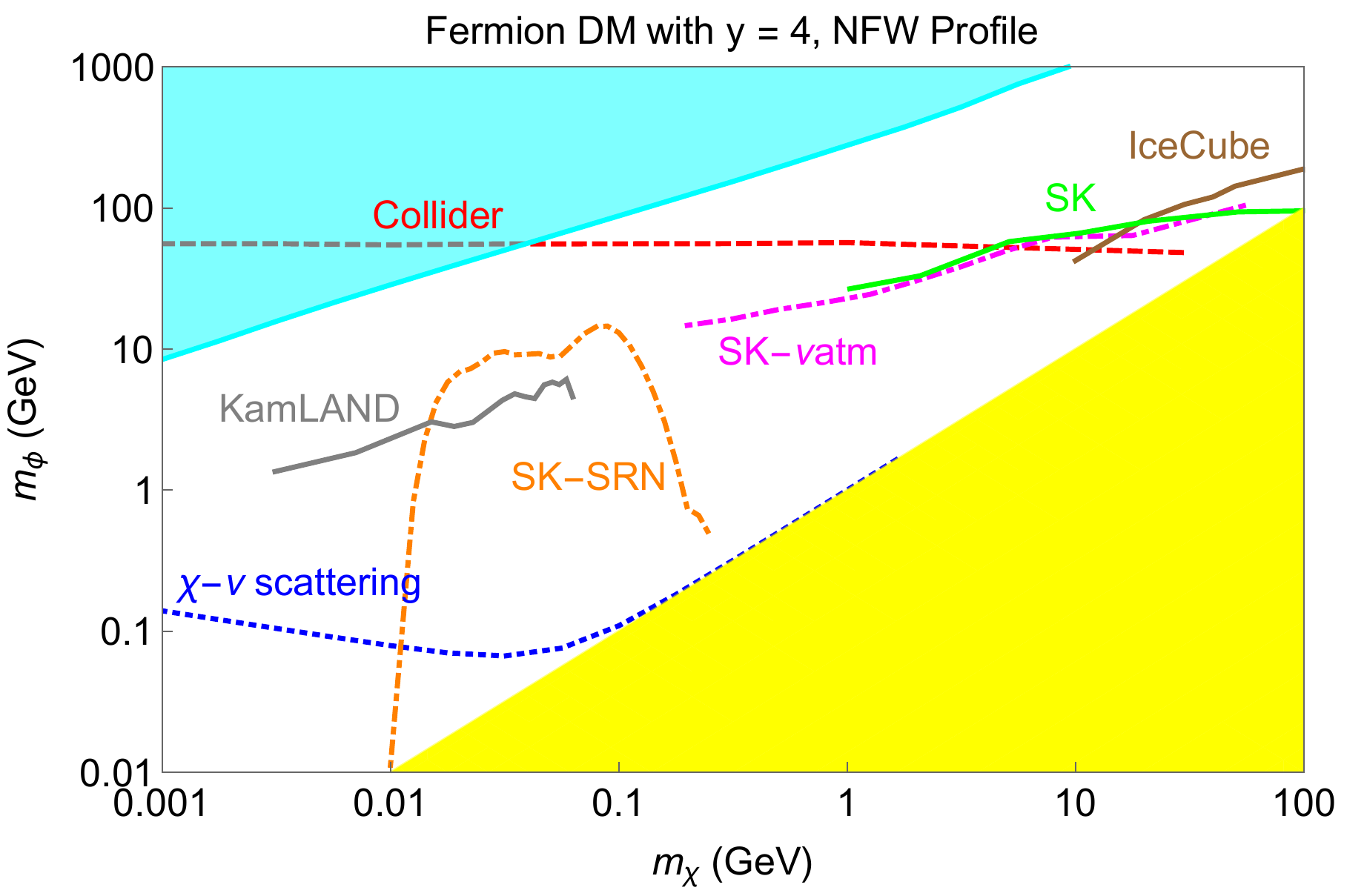}\label{fig:chiphi4}} 
        \caption{The 90\% CL bounds from various neutrino telescope observations compared with the collider bounds in the $m_\chi$-$m_\phi$ plane. The gray line, labelled KamLAND, is the (rescaled) official KamLAND bound~\cite{Collaboration:2011jza}. The green line, labelled SK, is the official SK bound~\cite{Frankiewicz:2015zma}. The brown line, labelled IceCube, is the official IceCube bound~\cite{Aartsen:2017ulx}. The orange, dot-dashed line, labelled SK-SRN, is the bound obtained from recasting the SK SRN searches discussed in Sec.~\ref{sk-snr}. The magenta, dot-dashed line, labelled SK-$\nu$atm, is the bound obtained from recasting the SK atmospheric neutrino measurements discussed in Sec.~\ref{sk-atm}. 
The blue dotted lines are the cosmological bounds from the DM-neutrino scattering discussed in Sec~\ref{sec:chinuscattering}. 
The dashed lines are the bounds obtained by taking the strongest bounds between the LHC mono-lepton and the invisible $Z$ decay width.
All the regions below the lines are excluded by the respective observations/experiments. The shaded cyan region is the region that gives $\Omega_\chi h^2 > 0.1186$ as simulated by Micromegas 5.0 \cite{Belanger:2018ccd}.  The shaded yellow region is the region which $m_\chi > m_\phi$. In these plots, the dark matter profile is taken to be the NFW profile. }
         \label{fig:mchimphi}
\end{figure}

\begin{figure}
       \centering
        \subfloat[$y = 1$]{\includegraphics[width= 0.49\textwidth]{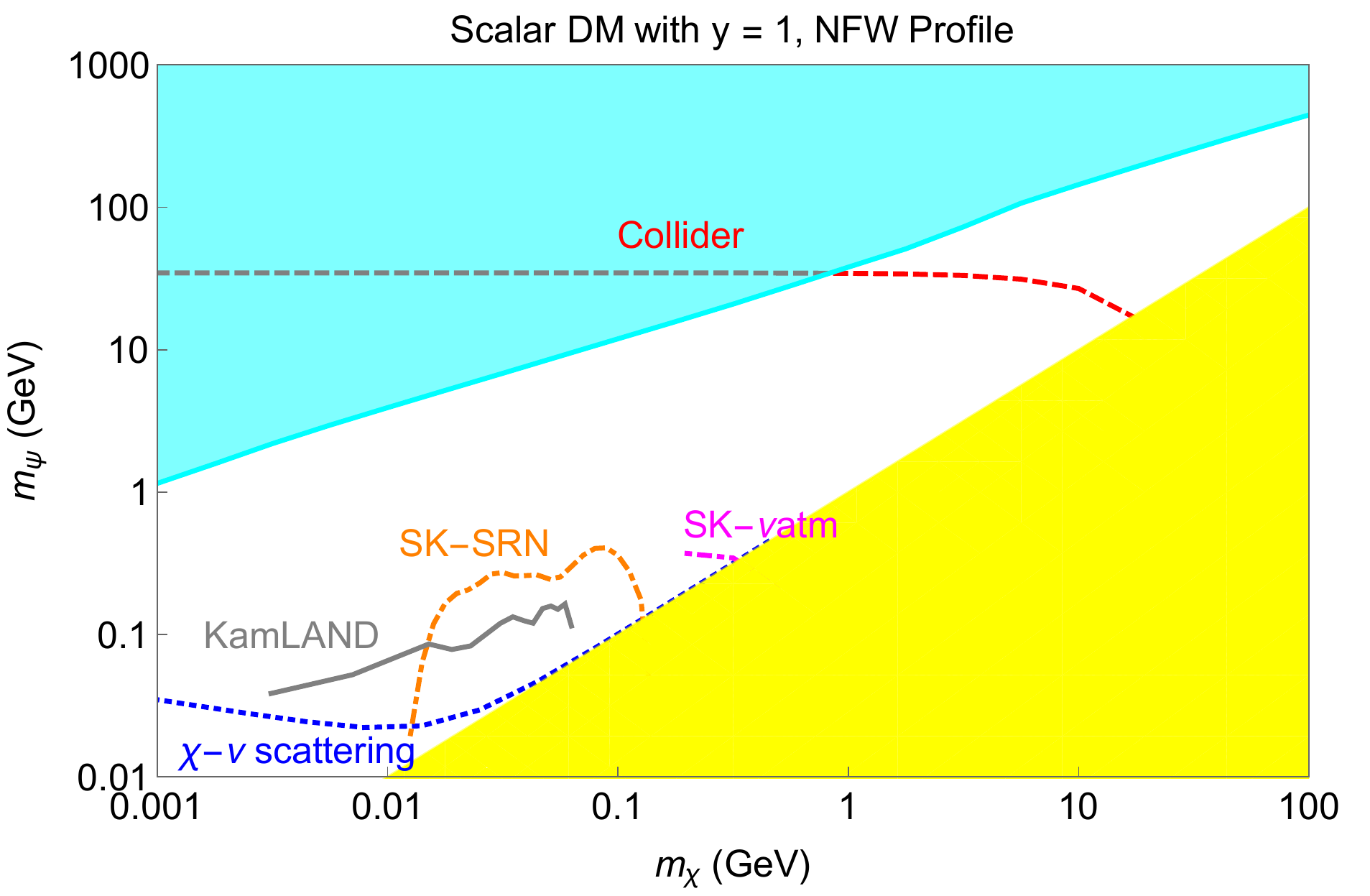}\label{fig:chipsi1}} 
        \subfloat[$y=4$]{\includegraphics[width= 0.49\textwidth]{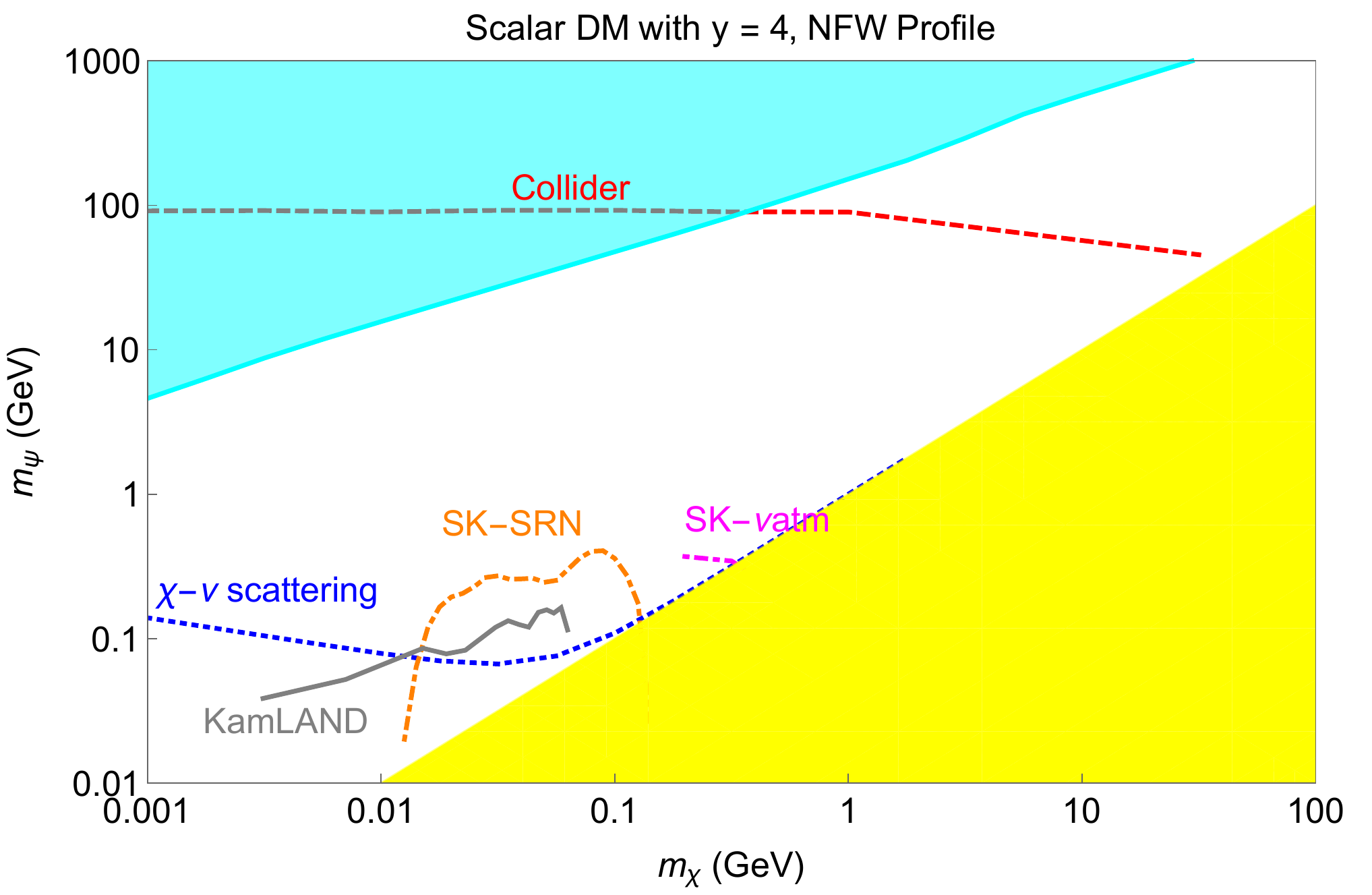}\label{fig:chipsi4}} 
       \caption{The 90\% CL bounds from various neutrino telescope observations compared with the collider bounds in the $m_\chi$-$m_\psi$ plane. The gray line, labelled KamLAND, is the (rescaled) official KamLAND bound~\cite{Collaboration:2011jza}. The green line, labelled SK, is the official SK bound~\cite{Frankiewicz:2015zma}. The brown line, labelled IceCube, is the official IceCube bound~\cite{Aartsen:2017ulx}. The orange, dot-dashed line, labelled SK-SRN, is the bound obtained from recasting the SK SRN searches discussed in Sec.~\ref{sk-snr}. The magenta, dot-dashed line, labelled SK-$\nu$atm, is the bound obtained from recasting the SK atmospheric neutrino measurements discussed in Sec.~\ref{sk-atm}. 
The blue dotted lines are the cosmological bounds from the DM-neutrino scattering discussed in Sec~\ref{sec:chinuscattering}. 
The dashed lines are the bounds obtained by taking the strongest bounds between the LHC mono-lepton and the invisible $Z$ decay width.
All the regions below the lines are excluded by the respective observations/experiments. The shaded cyan region is the region that gives $\Omega_\chi h^2 > 0.1186$ as simulated by Micromegas 5.0 \cite{Belanger:2018ccd}.  The shaded yellow region is the region which $m_\chi > m_\psi$. In these plots, the dark matter profile is taken to be the NFW profile. }
         \label{fig:mchimpsi}
\end{figure}

In this section we present all bounds obtained from the previous two subsections. Figures~\ref{fig:nfwbound} and~\ref{fig:burkertbound} show the bounds for both the fermionic and scalar DM cases in the $m_\chi$-$\langle\sigma v\rangle$ plane assuming the NFW and Burkert DM profile respectively. 
Figures~\ref{fig:mchimphi} and~\ref{fig:mchimpsi} display the same bounds in the $m_\chi$-$m_{\phi(\psi)}$ plane.
The bounds are compared against those obtained from the LHC and $Z$ decay width measurement in the context of simplified model of section~\ref{sec:framework}. Additionally, we also show the official bounds from IceCube~\cite{Aartsen:2017ulx}, Super-Kamiokande~\cite{Frankiewicz:2015zma} and KamLAND~\cite{Collaboration:2011jza} collaborations. Instead of picking a specific DM profile, the KamLAND collaboration uses a nominal value of $\mathcal J_{avg} = 5$. Thus, we rescale the KamLAND bounds using an appropriate $\mathcal J_{avg}$ factor of each DM profile. On the other hand, the Super-Kamiokande collaboration did the analysis for the NFW profile only. 
Hence, for the Burkert profile, we rescale the official SK bound with the appropriate $\mathcal J_{avg}$ factor.

From the plots we can see that in the case of the fermionic DM, the official bounds from SK and IceCube together with the bound from recast SK atmospheric neutrino analysis, which covers the mass region $m_\chi \gtrsim 200$ MeV, are stronger than the collider bounds. 
In the case of low dark matter mass with light mediator, the SK and KamLAND bounds are more constraining than the collider bounds. However, for low DM mass and heavy mediator, colliders provide stronger bounds.

For the case of scalar DM, its annihilation cross-section is p-wave suppressed. 
As a result, the collider bounds are more constraining than the neutrino telescope bounds for a low relative DM velocity.
In Figures~\ref{fig:nfwscalar},~\ref{fig:burkertscalar}, and~\ref{fig:mchimpsi} we show the bounds in the case of $v_{rel} = 10^{-3}$. 
For different value of $v_{rel}$, one can rescale the collider bound with a factor of $\left( \frac{v_{rel}}{10^{-3}}\right)^2$.
Note in particular that the collider bound in Figures~\ref{fig:nfwscalar} and \ref{fig:burkertscalar} for the case of $m_\psi=50$GeV is slightly stronger than the case of $m_\psi=100$GeV for low DM mass. This is because of the invisible Z decay bound gives a stronger constraint than a LHC bound.

\section{Comments on the Mediator and Possible UV Completions}
\label{sec:Mediator}

So far we have remained agnostic about the UV completion of the interaction in Eq.~\eqref{eq:lag}. 
In this section, we will briefly discuss the role of the mediator, in a UV complete model. 

It is possible to embed the interaction in Eq.~\eqref{eq:lag} in a UV complete model such that the mediator is an SM singlet. 
In this case, the DM ($\chi$), must be part of an electroweak doublet. For a concrete example, consider the supersymmetric Standard Model with right-handed neutrinos. We could have the DM being the neutralino while the mediator ($\phi$) is the right-handed sneutrino.  

Another possible UV completion is to have the mediator being a neutral component of an electroweak doublet. In this case, we would have the DM being an SM singlet. For the case of a fermionic DM, the scalar mediator resembles the a slepton of the supersymmetric Standard Model. Hence this scenario is constrained by the LHC slepton searches. The latest LHC result puts the lower bound on the mass of charged partner of the mediator ($m_{\phi^\pm}$) to be around 500 GeV for $m_{\phi^\pm} - m_{\chi} \gtrsim 60$ GeV~\cite{ATLAS:2017uun}. Since we consider the case of $m_\phi \lesssim 100$ GeV, to satisfy this bound, the mass of the neutral and charged component of the doublet must be split. This could be achieved by introducing an interaction with the Higgs doublet such as $\mathcal L \supset \lambda_{H\Phi} (H^\dagger \Phi)(\Phi^\dagger H)$, where $\Phi$ is the electroweak doublet containing the mediator. Alternatively, one can avoid the bound on the mass of the charged partner of the mediator in the case of the compressed spectra, $m_{\phi^\pm} \sim m_{\phi} \lesssim m_\chi + 60$ GeV. In this case the leptons coming from the decay of the charged partner will be too soft to pass the search cuts.

We could instead have a fermionic mediator being part of an electroweak doublet, $\Psi$. The mediator must be vector-like to avoid anomalies, ie.
\begin{equation}
\mathcal L \supset y \bar \Psi_R L_L \chi + M \bar\Psi_R \Psi_L+ \text{ h.c.},
\end{equation}
where $L_L$ is the left-handed lepton doublet. $\Psi_L$ and $\Psi_R$ are the left- and right-handed fermion doublets with the same charge assignment as the SM left-handed lepton doublet. 
As in the scalar case, there exists a bound from the charged partner of the mediator. The current LHC bounds on a vector-like lepton is $m_\psi > 176$ GeV~\cite{Aad:2015dha}.

Lastly, both the DM and the mediator could be singlets under the SM gauge groups. The dark sector communicates with the SM sector via mixing with a right handed neutrino of the see-saw mechanism. Several bounds regarding the right handed neutrino in this scenario are discussed in \cite{Batell:2017rol, Batell:2017cmf}.

\section{Conclusions and discussions}
\label{sec:conclusion}
DM interactions are among one of the least understood interactions in nature. There is hope that the LHC can provide insight into DM interactions.   

In this work, we consider the case where DM only interact with the SM through a neutrino and a mediator.  
We study the interaction in the framework of the simplified model, see Eq.~\eqref{eq:lag}. 
In this scenario, there are two viable options to probe the DM interaction. 
First, one can probe such an interaction using the mono-lepton and the mono-jet signal at the LHC. 
These experimental signatures arise from the $W$ and $Z$ production.
The $W$ and $Z$ decays, via the off-shell neutrino, to DM and the mediator as shown in Fig.~\ref{fig:3bodydecay}.
By reinterpreting ATLAS mono-lepton result~\cite{Aaboud:2017efa} as well as the invisible $Z$ width measurement~\cite{Olive:2016xmw}, we derive the bounds on DM-neutrino coupling shown in Fig.~\ref{fig:ybound}.

Moreover, one can probe the DM-neutrino interaction using the neutrino signal coming from DM annihilation in the galaxy.
Here, we consider both the official bounds from a dedicated neutrino from DM annihilation search at IceCube~\cite{Aartsen:2016pfc,Aartsen:2017ulx}, KamLAND~\cite{Collaboration:2011jza} and SK~\cite{Frankiewicz:2015zma}, as well as our reinterpretation of SK data from SRN search, see Sec.~\ref{sk-snr}, and the atmospheric neutrino measurement, see Sec.~\ref{sk-atm}.

We found that in the case of fermionic DM, neutrino telescopes give the best constraint for $m_\chi \gtrsim 200$ MeV. For lower DM mass, neutrino telescopes continue to provide the best constraint for a light mediator. However, for low DM mass with a heavy mediator, the LHC provides the better bounds, see Fig.~\ref{fig:nfwfermion} and~\ref{fig:burkertfermion}.
For the case of scalar DM, we found that the LHC bounds are generally more stringent than the bound from the neutrino telescope for all DM mass considered in this work, i.e. $1 \text{ MeV}< m_\chi < 100$ GeV, see Fig.~\ref{fig:nfwscalar} and~\ref{fig:burkertscalar}. This can be understood by noting that the DM annihilation cross-section for the scalar DM is velocity suppressed as shown in Eq.~\eqref{eq:xsec}. Hence the LHC bound on DM coupling, when interpreted in term of the bound on DM annihilation cross-section, get suppressed by a small relative DM velocity.
\begin{acknowledgments}
The work of PU has been supported in part by 
the Thailand Research Fund under contract no.~MRG6080290, and the Faculty of Science, Srinakharinwirot University under grant no.~422/2560. We thank Joachim Kopp for his valuable comments on our manuscript. We also would like to thank the referees for their useful comments and suggestions.

\end{acknowledgments}

\bibliography{lit1} \bibliographystyle{JHEP}

\end{document}